\documentclass[
 reprint,
 amsmath,
 amssymb,
 aps,
]{revtex4-2}

\usepackage[hyperref]{xcolor}

\definecolor{goethe-blau}{cmyk}{1.0,0.2,0.0,0.4}
\definecolor{hellgrau}{cmyk}{0.04,0.04,0.05,0.02}
\definecolor{sandgrau}{cmyk}{0.12,0.09,0.13,0.0}
\definecolor{dunkelgrau}{cmyk}{0.25,0.25,0.30,0.75}
\definecolor{emo-rot}{cmyk}{0.04,1.0,0.8,0.07}
\definecolor{purple}{cmyk}{0.08,1.0,0.3,0.36}
\definecolor{senfgelb}{cmyk}{0.01,0.25,1.0,0.05}
\definecolor{gruen}{cmyk}{0.62,0.4,0.87,0.09}
\definecolor{magenta}{cmyk}{0.08,0.86,0.12,0.12}
\definecolor{orange}{cmyk}{0.0,0.7,1.0,0.04}
\definecolor{sonnengelb}{cmyk}{0.0,0.12,0.95,0.0}
\definecolor{helles-gruen}{cmyk}{0.4,0.17,0.81,0.07}
\definecolor{lichtblau}{cmyk}{0.8,0.0,0.06,0.04}

\usepackage[
 colorlinks,
 pdfpagelabels,
 breaklinks,
 pdfstartview=FitH,
 bookmarksopen=true,
 bookmarksnumbered=true,
 bookmarksopenlevel=2,
 plainpages=false,
 hypertexnames=false,
 citecolor=emo-rot,
 linkcolor=goethe-blau,
 urlcolor=purple,
 pdftitle={},
 pdfauthor={},
 pdfkeywords={}
]{hyperref}

\usepackage[capitalise,sort,compress]{cleveref}

\Crefname{section}{Sec.}{Sections}
\Crefname{table}{Tab.}{Tables}

\usepackage{comment}

\usepackage[acronym]{glossaries-extra}

\glssetcategoryattribute{acronym}{nohyperfirst}{true}
\setabbreviationstyle[acronym]{long-short}

\usepackage{bm}
\usepackage{dsfont}
\usepackage{tensor}
\usepackage{amsthm}
\usepackage{physics}
\usepackage{upgreek}
\usepackage[bb=boondox]{mathalfa}
\usepackage{mathtools}

\usepackage{ulem}

\usepackage{graphicx}
\usepackage{dcolumn}
\usepackage{float}
\usepackage[percent]{overpic}

\newacronym{MFA}{MFA}{mean-field approximation}
\newacronym{NJL}{NJL}{Nambu-Jona-Lasinio model}
\newacronym{SSB}{SSB}{spontaneous symmetry breaking}
\newacronym{CSB}{CSB}{chiral symmetry breaking}
\newacronym{QMM}{QMM}{quark-meson model}
\newacronym{QCD}{QCD}{quantum chromodynamics}
\newacronym{HIC}{HIC}{heavy ion collisions}

\begin{document}


\preprint{APS/123-QED}

\title{Magnetization by Rotation: Spin and Chiral Condensates in the NJL Model}

\author{Lutz Kiefer}
\affiliation{Institut f\"ur Theoretische Physik, Goethe University,
Max-von-Laue-Straße 1, D-60438 Frankfurt am Main, Germany}

\author{Ashutosh Dash}
\affiliation{Institut f\"ur Theoretische Physik, Goethe University,
Max-von-Laue-Straße 1, D-60438 Frankfurt am Main, Germany}

\author{Dirk H.~Rischke}
\affiliation{Institut f\"ur Theoretische Physik, Goethe University,
Max-von-Laue-Straße 1, D-60438 Frankfurt am Main, Germany}
\affiliation{
	Helmholtz Research Academy Hesse for FAIR, Campus Riedberg, Max-von-Laue-Straße 12, D-60438 Frankfurt am Main, Germany
}

\date{\today}

\begin{abstract}
The role of spin degrees of freedom in the quark-gluon plasma (QGP) has attracted significant interest in recent years. 
Spin hydrodynamics extends conventional hydrodynamics by incorporating spin via the spin tensor. 
In the mean-field limit of the Nambu–Jona-Lasinio (NJL) model under rigid rotation, spin degrees of freedom manifest naturally as axial-vector, or spin, condensate. 
We investigate the interplay between chiral and spin condensates in this framework. 
While rotation typically suppresses the formation of a chiral condensate, the presence of a spin condensate may counteract this effect, enhancing the chiral condensate.
Moreover, it can alter the nature of the chiral transition from second  to first order.
\end{abstract}

\maketitle


\textit{Introduction}: 
In the context of relativistic heavy-ion collisions, where polarization effects have been experimentally observed \cite{STAR:2017ckg}, the role of spin degrees of freedom has attracted significant attention in recent years \cite{Becattini:2022zvf,Becattini:2011ev,Becattini:2018duy,Becattini:2020riu,Buzzegoli:2021wlg,Speranza:2020ilk,Fukushima:2020ucl,Das:2021aar,Dey:2023hft,Florkowski:2018fap,Li:2020eon,She:2021lhe,Biswas:2023qsw,Florkowski:2017ruc,Florkowski:2017dyn,Florkowski:2018myy,Weickgenannt:2019dks,Bhadury:2020puc,Weickgenannt:2020aaf,Shi:2020htn,Bhadury:2020cop,Singh:2020rht,Peng:2021ago,Sheng:2021kfc,Weickgenannt:2022qvh,Gallegos:2020otk,Garbiso:2020puw,Montenegro:2017rbu,Montenegro:2017lvf,Liu:2021nyg,Hongo:2021ona,Weickgenannt:2022zxs}. 
Beyond high-energy physics, spin degrees of freedom also play a crucial role in astrophysical settings such as neutron stars, where extremely strong magnetic fields are believed to originate, in part, from the spin polarization of dense nuclear matter \cite{Lattimer:2004pg}. 
Moreover, a consistent description of spin dynamics is not limited to nuclear or astrophysical systems; it is also fundamental in condensed-matter physics, with important applications in magnetic materials and spintronics (see, e.g., the recent reviews \cite{RevModPhys.76.323,RevModPhys.87.1213}).

From a theoretical point of view, hydrodynamics provides a universal  effective low-energy description of many-body systems. 
When generalized to include spin, the conservation of total angular momentum leads to a (non-)conservation equation for the spin current, which naturally introduces a spin potential -- the thermodynamic quantity conjugate to the spin density, analogous to a chemical potential for spin. 
In the context of the formation of the quark-gluon plasma (QGP), such a framework -- often referred to as spin hydrodynamics -- may be essential for understanding spin-related observables reported in experiments \cite{STAR:2017ckg}. 
One of the dynamical variables of spin hydrodynamics is the spin tensor. 
For a system of Dirac fermions, the canonical spin tensor can be obtained directly from Noether’s theorem, 
\begin{align}\label{Eq:Canonical_ST}
    S_C^{\lambda , \mu \nu} &= \frac{1}{2} \bar{\psi} \left\lbrace \gamma^\lambda \, , \, \Sigma^{\mu\nu}\right\rbrace \psi = \frac{1}{2} \epsilon^{\rho\lambda\mu\nu} j_{5, \rho} \;,
\end{align}
where $\Sigma^{\mu\nu}=i/4 \left[ \gamma^\mu \, ,\,\gamma^\nu \right]\,$, is the so called spin-density matrix and
$j_5^\mu =\bar{\psi} \gamma^\mu \gamma^5 \psi$ is the axial-vector current. 
The canonical spin tensor is completely anti-symmetric and therefore only has four independent components. 
Hence, the canonical spin tensor is in one-to-one correspondence  to the axial-vector current $j_5^\mu$. 
The above relation will be crucial for the remainder of this paper.

Within the framework of hydrodynamics, there is an inherent ambiguity in the definition of the spin tensor, reflecting the freedom allowed by pseudogauge transformations \cite{Hongo:2021ona,Gallegos:2020otk}. 
This ambiguity is a generic feature of effective theories, where multiple definitions can yield the same macroscopic conservation laws.
For the sake of definiteness, we choose the canonical spin tensor \eqref{Eq:Canonical_ST} in this work.

In this paper, our primary interest lies not in spin hydrodynamics, but rather in understanding the thermodynamic properties of matter under rotation and the possible phases associated with spin degrees of freedom. 
In particular, we explore the emergence of spin polarization and potential ferromagnetic phases in rotating quark matter. 
This idea draws an interesting parallel to a well-known effect observed over a century ago -- the Barnett effect -- where an uncharged rotating body becomes magnetized, due to the alignment of intrinsic spins along the axis of rotation \cite{PhysRev.6.239}. 

The possibility of ferromagnetism in an electron gas was first proposed in 1929 within QED \cite{bloch1929bemerkung} and later confirmed via Monte Carlo methods \cite{PhysRevLett.45.566}.
At low densities, the Coulomb exchange interaction becomes dominant over the kinetic energy, leading to a first-order phase transition from an unpolarized to a fully polarized state. 
In high-density QCD, similar spin-polarized phases have been explored, particularly in the context of color superconductivity and within  Nambu--Jona-Lasinio (NJL) models at zero temperature and moderate baryon density \cite{Tsue:2012nx, Tsue:2014tra, Matsuoka:2016ufr, Maedan:2006ib, Tsue:2012jz, Matsuoka:2016spx, Morimoto:2018pzk}. 

In this paper, we model the dynamical generation of magnetization by introducing an axial-vector interaction in the NJL-model Lagrangian \cite{PhysRev.122.345,Klevansky1992},
\begin{align}\label{eq:NJL_model}
    \mathcal{L}_{\text{NJL}} = \bar{\psi}(i\gamma^\mu \partial _\mu -m) \psi + G (\bar{\psi}\psi)^2 + G_A (\bar{\psi}\gamma_\mu \gamma^5\psi)^2  \,.
\end{align}
The first term in the Lagrangian represents the free Dirac Lagrangian of fermions with constituent mass $m$, the second corresponds to a scalar  interaction, and the third describes an axial-vector current interaction, i.e., due to Eq.~(\ref{Eq:Canonical_ST}) essentially a spin-spin interaction. 
In this work, for the sake of simplicity we only adopt the scalar interaction (coupling constant $G$) and the spin-spin interaction (coupling constant $G_A$) as an effective description of quark spin interactions in QCD at nonzero temperature and density. 
Notably, a recent study \cite{Buzzegoli_2024} has established a connection between spin-spin interactions in the presence of rotation and the emergence of the canonical spin tensor within the framework of ideal spin hydrodynamics.\\

\textit{NJL model under rigid rotation}: In a nontrivial, i.e., rotating, background the kinetic term in the Lagrangian (\ref{eq:NJL_model}) has to be replaced by
\begin{equation}\label{eq:NJL_model_rotating}
\bar{\psi}i\gamma^\mu \partial_\mu \psi \quad \longrightarrow  \quad \bar{\psi}i\gamma^a D_a \psi \;,
\end{equation}
where $\gamma^a$ are the 4×4 Dirac matrices in flat spacetime, satisfying the anticommutation relation
\begin{equation}
\left\lbrace \gamma^a , \gamma^b \right\rbrace = 2\eta^{ab} ~,
\end{equation}
where $\eta^{ab} = \text{diag}(+,-,-,-)$ is the metric tensor in flat spacetime.
The rotating background appears in the covariant derivative, which is defined as
\begin{equation}
D_a = e_a^{\mu} \left(\partial_\mu - \Gamma_\mu \right)~.
\end{equation}
where Latin indices $a,b, \ldots$ label the local (tetrad) frame, and Greek indices $\mu, \nu, \ldots$ denote spacetime coordinates. 
The fields $e^a_\mu$ are the tetrads, with $e^a_\mu e_b^\mu = \delta^{a}_b$, and $\Gamma_\mu$ are the spin-connection terms arising from the nontrivial (rotating) background, given by
\begin{equation}\label{Gamma_mu}
\Gamma_\mu = -\frac{1}{4} \gamma^a \gamma^b e_a^\nu \partial_\mu e_{b\nu} ~.
\end{equation}
To study the system under rotation, we adopt the metric of a cylinder rigidly rotating with constant angular velocity $\Omega$:
\begin{equation}
ds^2 = \left(1 - \Omega^2 r^2\right) dt^2 - 2 \Omega r^2 d\theta dt - dr^2 - r^2 d\theta^2 - dz^2 ~.
\end{equation}
We choose the nonvanishing components of the tetrad fields as
\begin{align}
e_0^t &= 1 ~, & e_1^r &= 1 ~, & e_3^z &= 1 ~, &\
e_0^{\theta} &= -\Omega ~, & e_2^{\theta} &= \frac{1}{r} ~ .
\end{align}
Substituting this choice into \cref{Gamma_mu}, the nonzero components of the spin connection become
\begin{equation}
\Gamma_t = \Omega \Gamma_{\theta} ~, \quad \Gamma_{\theta} = -\frac{1}{2} \gamma^1 \gamma^2 ~, \quad \Gamma_r = \Gamma_z = 0 ~.
\end{equation}
\\
\textit{Effective potential in the mean-field approximation}: We perform a Hubbard-Stratonovich transformation to replace the four-fermion interaction terms in the NJL Lagrangian \eqref{eq:NJL_model} in terms of scalar and axial-vector fields. and then take the mean-field limit, i.e., approximate the latter by their saddle-point values,
\begin{align}
\langle \bar{\psi} \psi \rangle &= -\frac{\sigma - m}{G}\; , \quad \langle \bar{\psi}\gamma^\mu \gamma^5\psi \rangle = -\frac{s^\mu}{G_A} \;, 
\end{align} 
where $\sigma$ is the chiral condensate and $s^\mu$ is the spin condensate. 

The mean-field partition function reads:
\begin{align}
Z = \int \mathcal{D}\bar{[\psi]}\mathcal{D}[\psi] \,\exp\left(i S\left[ \sigma, s^\mu \right] \right) \, ,
\end{align}
where the action $S$ is given by
\begin{align}
    S\left[ \sigma, s^\mu \right] &=\int d^4x \sqrt{g}\, \bar{\psi}(x) \left( i \gamma^a D_a - \sigma  - \gamma^3 \gamma^5 |\mathbf{s}| \right) \psi(x)\nonumber\\
    & - \frac{V_4}{2G} \left(\sigma - m \right) ^2 - \frac{V_4}{2 G_A} \mathbf{s}^2\,,
\end{align}
with  $g = \det(e_a^\mu)$ and $V_4$ denoting the spacetime volume of the system.
Without loss of generality, we choose the components of the spin condensate $s^\mu$ orthogonal to the rotation axis (i.e., the $z$-axis) to be zero, i.e., $s^\mu = (0,0,0,|\mathbf{s}|)$.

To compute the effective potential, we employ the Ritus method and perform the Matsubara sum. 
The details of the calculation are provided in the supplementary material. 
For an infinite system, the transverse momentum is a continuous variable and the final expression for the effective potential is given by
\begin{widetext}
\begin{align}\label{eq:Potential_final}
    V_{\text{eff}}&=\frac{(\sigma-m)^2}{2G}+\frac{\mathbf{s}^2}{2G_A}-\frac{N_f N_c}{(2\pi)^3}\sum_{l=0}^\infty\sum_{e=\pm} \sum_{s=\pm}\int p_\bot dp_\bot dp_z  \bigg[ T\,\ln(1+e^{-(E_s-e\Omega l)/T}) 
    +E_s\bigg]\,,
\end{align}
where $e=\pm$ denotes the contribution of particles/antiparticles, $s=\pm$ denotes spin up/down, and
\begin{align}\label{Eq:DispersnRelan}
    E_{s} = \sqrt{p_\perp^2 + \left( \sqrt{p_z^2+\sigma^2}+ s|\mathbf{s}|\right)^2}
\end{align}
\end{widetext}
is the dispersion relation for particles with momentum $\mathbf{p}$ and spin direction $s$.
\begin{figure*}
\centering

\begin{overpic}[width=0.23\textwidth]{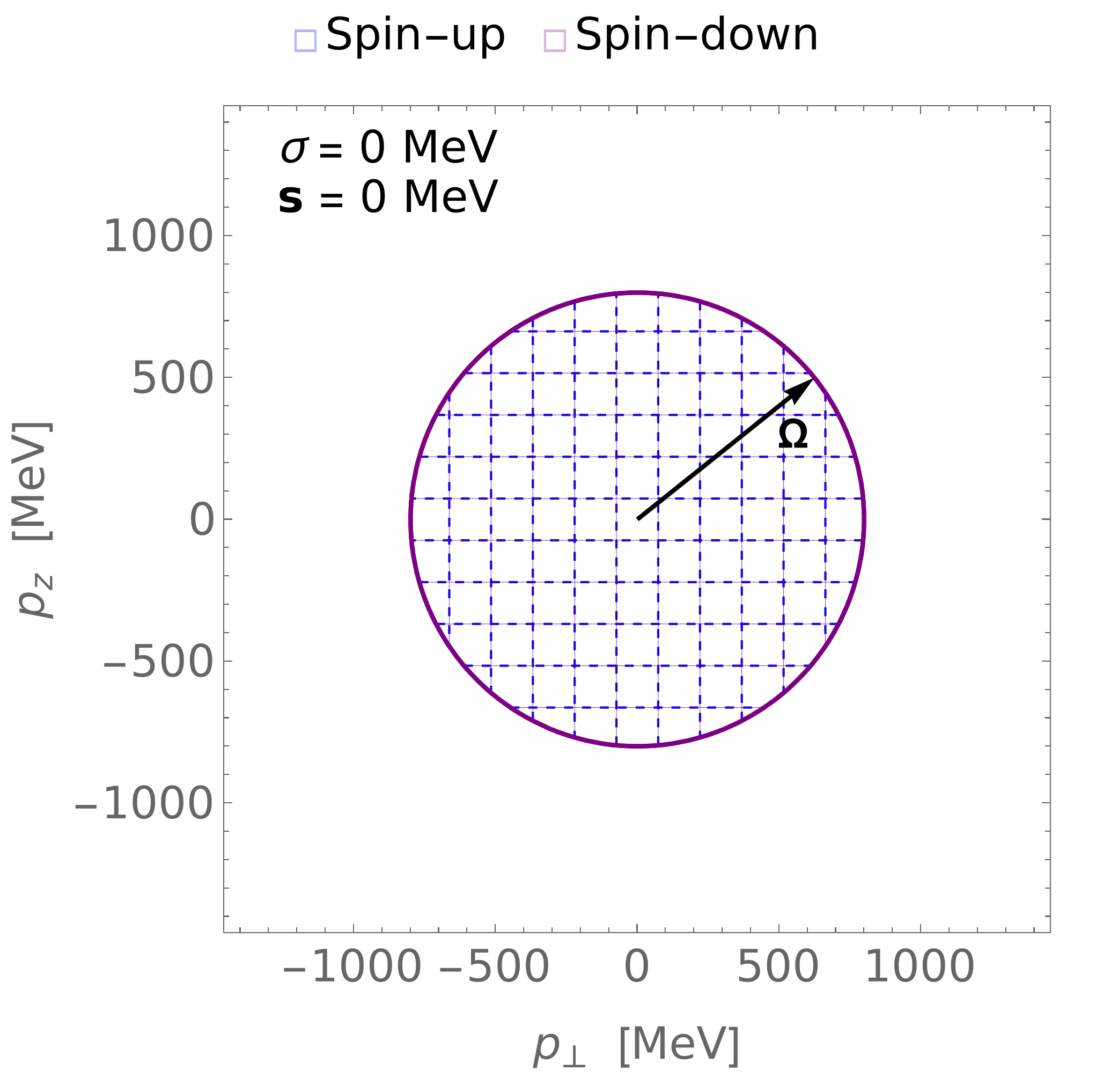}
    \put(5,90){(a)}
\end{overpic}
\hfill
\begin{overpic}[width=0.23\textwidth]{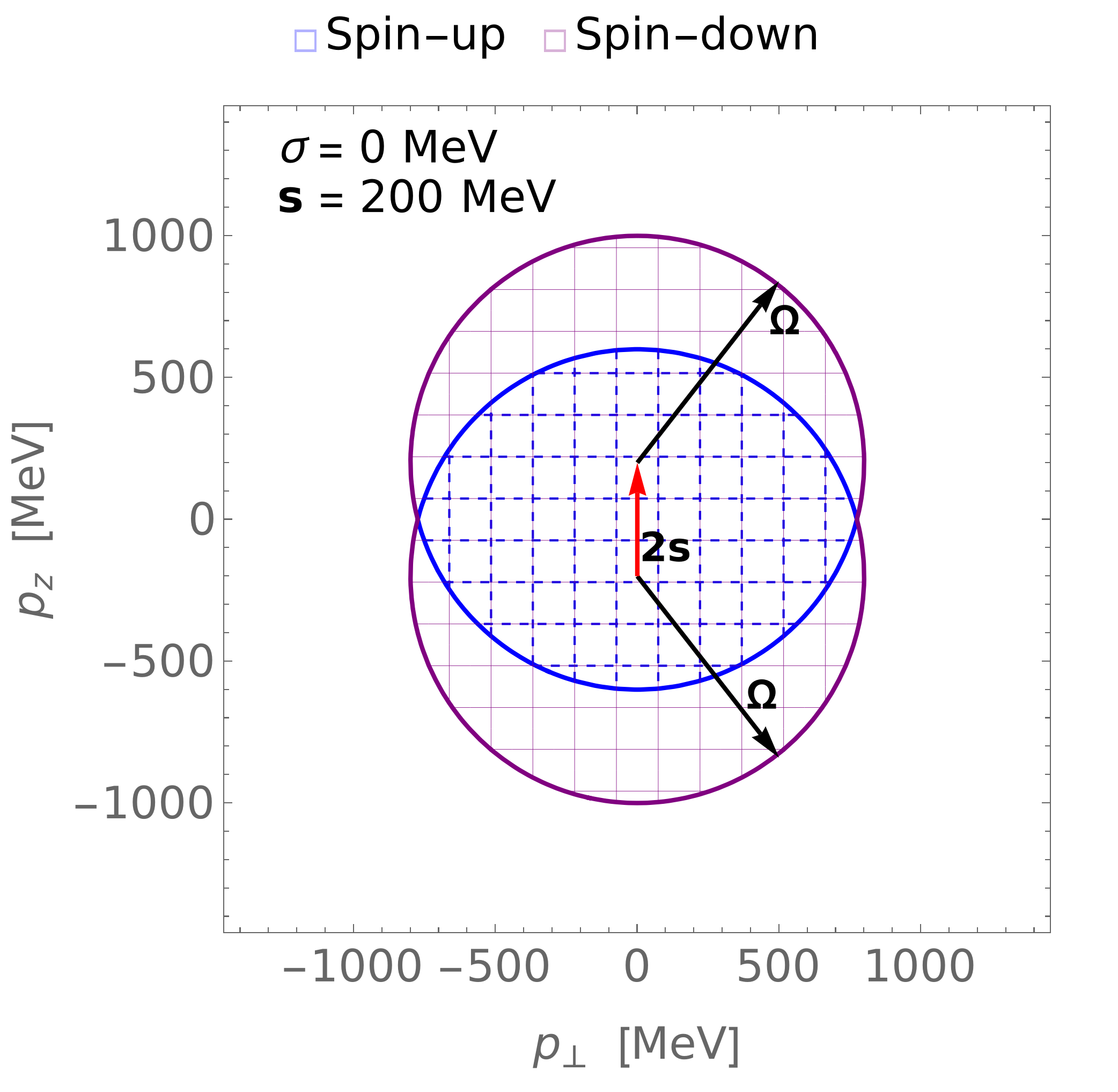}
    \put(5,90){(b)}
\end{overpic}
\hfill
\begin{overpic}[width=0.23\textwidth]{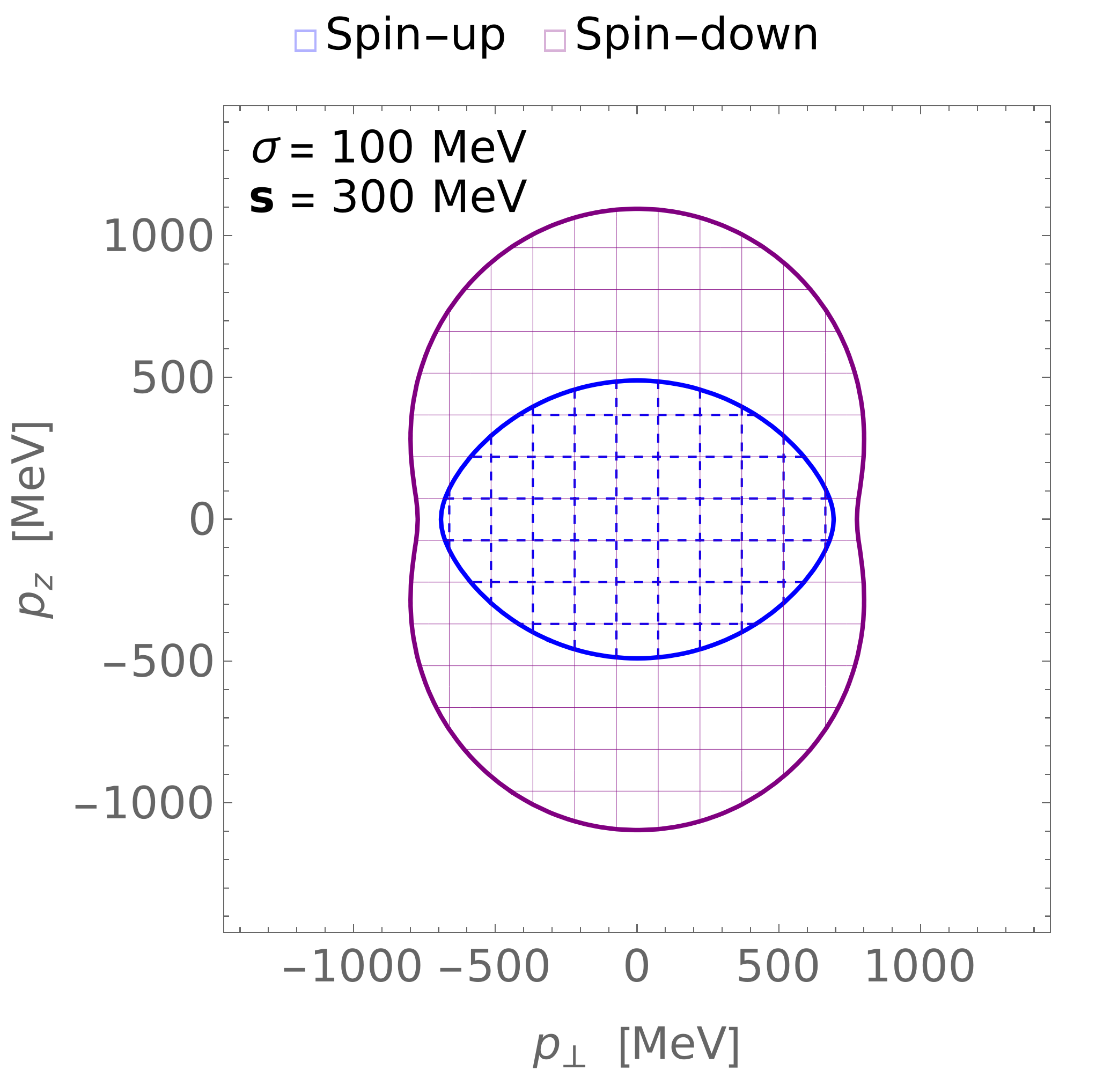}
    \put(5,90){(c)}
\end{overpic}
\hfill
\begin{overpic}[width=0.23\textwidth]{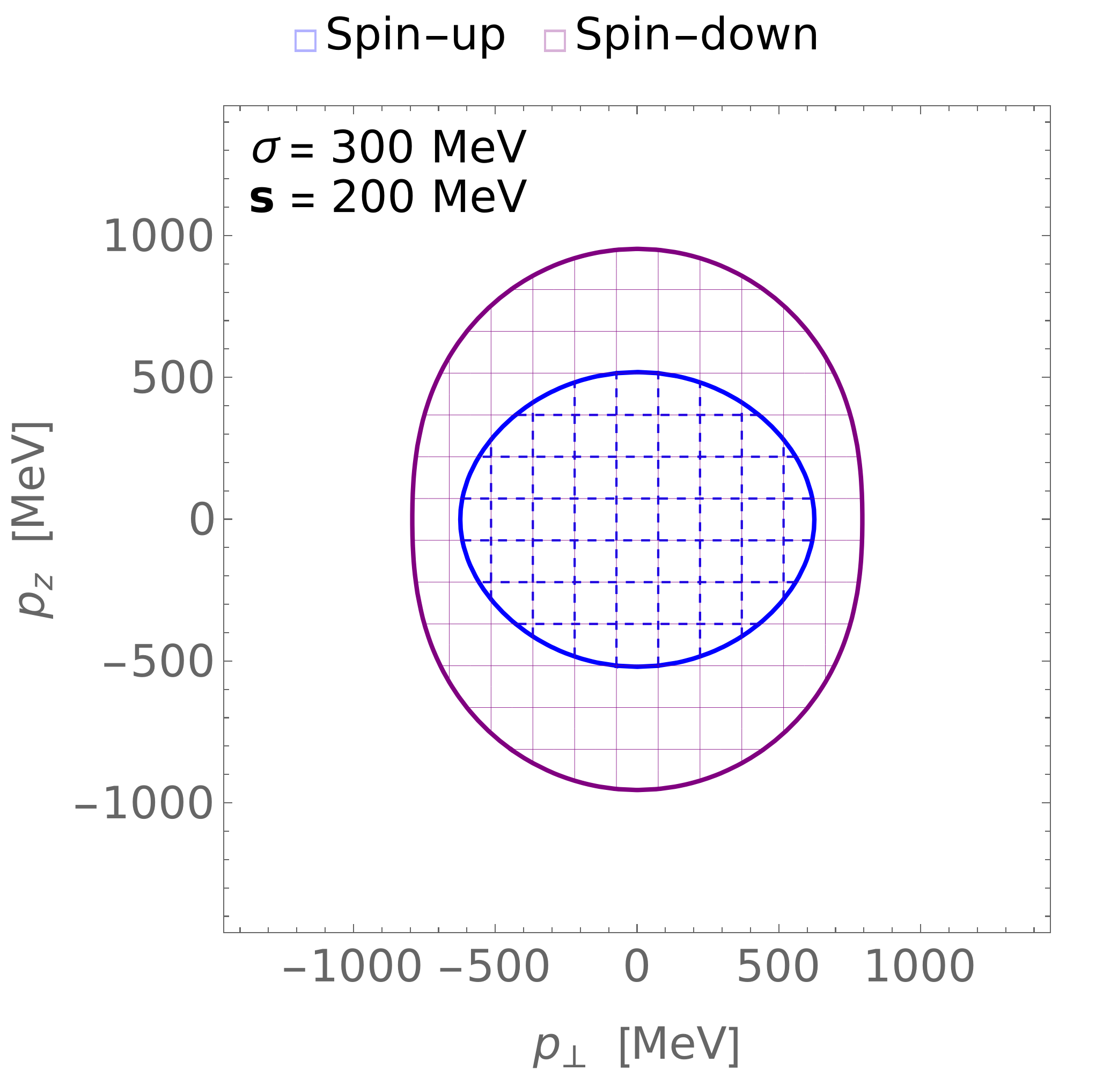}
    \put(5,90){(d)}
\end{overpic}

\caption{\label{Fig:Fermi_surface}
Fermi-sea profiles in the \(p_z\)–\(p_\perp\) plane for \(\Omega = 800\,\mathrm{MeV}\).
The panels correspond to:
(a) \(\sigma = 0\), \(\mathbf{s} = 0\);
(b) \(\sigma = 0\), \(|\mathbf{s}| = 200\,\mathrm{MeV}\);
(c) \(\sigma = 100\,\mathrm{MeV}\), \(|\mathbf{s}| = 300\,\mathrm{MeV}\);
(d) \(\sigma = 300\,\mathrm{MeV}\), \(|\mathbf{s}| = 200\,\mathrm{MeV}\).
Blue and purple regions indicate spin-up and spin-down Fermi seas, respectively.
All cases exhibit axial and reflection symmetry.
A nonzero \(|\mathbf{s}|\) leads to deformation and splitting of the Fermi surfaces.}
\end{figure*}
The factor $s=\pm$ in front of the spin condensate $|\mathbf{s}|$ indicates the energy splitting between different spin states due to the presence of the spin condensate, which corresponds to the exchange splitting in a non-relativistic electron system \cite{yosida1996theory}. 
The presence of the spin condensate introduces an anisotropy in the momentum-space integral in \cref{eq:Potential_final}.
We note that causality requires an upper bound on the radial extent of the rotating cylinder. 
For the moment, we neglect this, but we will discuss the implications of this constraint below.

We use the Pauli-Villars regularization scheme \cite{Pauli_Villars} to render the vacuum contribution in \cref{eq:Potential_final} finite, i.e., we replace $E_s$ the last term in \cref{eq:Potential_final} by
\begin{align}
E_s \quad \longrightarrow \quad  \sum_{j=0}^3 c_j \sqrt{E_s^2+j\Lambda^2}\, .
\end{align}
Here we choose $c_0=1\, , \,c_1=-3\, , \,c_2=3\, , \,c_3=-1 $, and  $\Lambda=757.048 ~\, \text{MeV}$.\\

\textit{Influence of the spin condensate on the Fermi sea}: Prior to finding the minimum of the effective potential and solving the respective equations self-consistently, it is instructive to examine the structure of the Fermi sea in the presence of the spin condensate. 
Notably, the vacuum contribution to \cref{eq:Potential_final} is independent of the angular velocity $\Omega$; thus, any rotational effects arise solely from the medium-dependent part of the potential.

In the zero-temperature limit, the following identity is useful:
\begin{equation}
\lim_{T \to 0} T \ln\left(1 + e^{-(E - \mu)/T}\right) = (\mu - E)\, \theta(\mu - E) \, ,
\end{equation}
which allows us to evaluate the medium contribution to \cref{eq:Potential_final} as $T\to 0$. 

In \cref{Fig:Fermi_surface}, we display the Fermi seas for particles ($e=+$) with spin direction $s=\pm$, projected onto the $p_z-p_\bot$ plane for the following cases: (a) $\sigma=0,~|\mathbf{s}|=0$, (b) $\sigma=0,~|\mathbf{s}|>0$, (c) $\sigma\neq 0,~|\mathbf{s}|>\sigma$, and (d) $\mathbf{s}\neq0,~|\mathbf{s}|<\sigma$. 
For the sake of simplicity, we only consider the $l=1$ term in \cref{eq:Potential_final}.
In all four scenarios, the Fermi seas exhibit axial symmetry about the $z$-axis and reflection symmetry with respect to the transverse plane. 
The purple shaded region represents the Fermi sea of spin-down quarks, while the blue shaded region corresponds to spin-up quarks. 
From the resulting profiles, we observe:
\begin{enumerate}
    \item[(a)] For $\sigma=|\mathbf{s}|=0$, the spin-up and spin-down states occupy identical circular regions in momentum space, each with radius $\Omega$, resulting in symmetric Fermi surfaces.
    \item[(b)] For \(\sigma = 0\), \(|\mathbf{s}| = 200\, \text{MeV}\), the total occupied momentum-space volume remains unchanged from case (a), however, spin-up particles occupy a smaller volume than spin-down particles, indicating a net spin polarization in the system. 
    \item[(c)] For \(\sigma = 100\, \text{MeV},~|\mathbf{s}| = 300\, \text{MeV}\), the Fermi surfaces deform and separate. The spin-down Fermi sea expands while the spin-up sea contracts.
    \item[(d)] For \(\sigma = 300\,\text{MeV},~|\mathbf{s}| = 200\, \text{MeV}\), the Fermi surfaces for spin-down and spin-up states are deformed into prolate and oblate shapes, respectively.  
\end{enumerate}

\begin{figure*}
\includegraphics[scale=0.3]{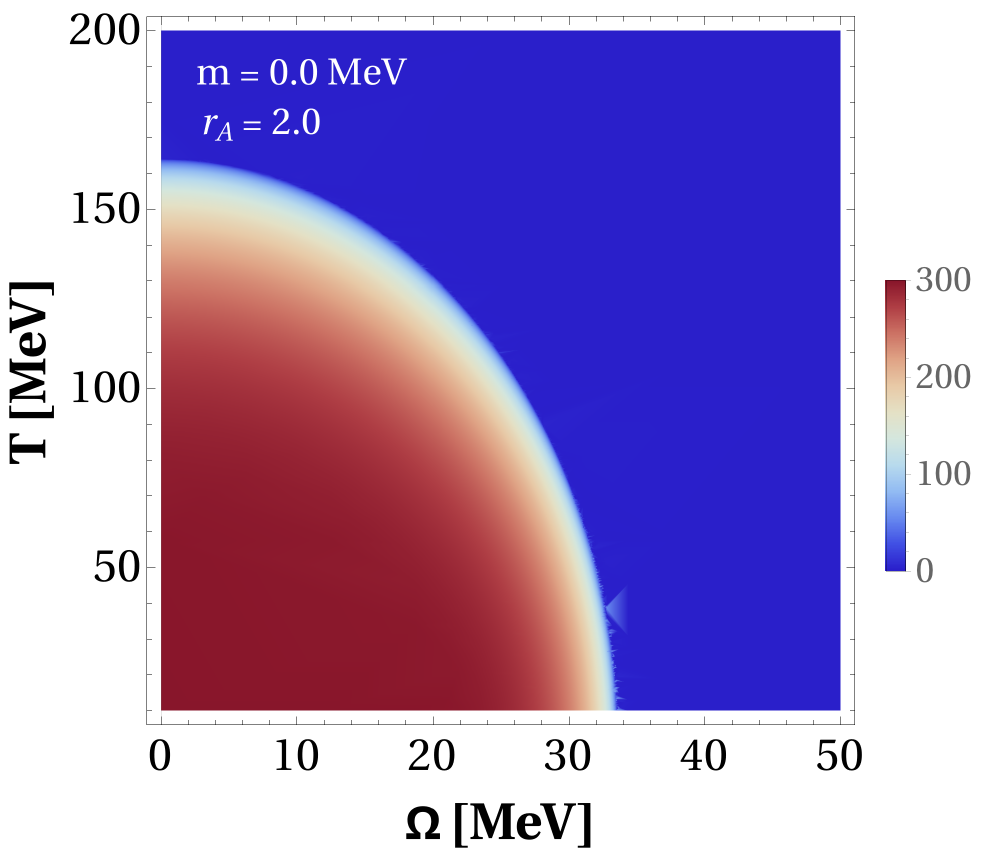}
\includegraphics[scale=0.3]{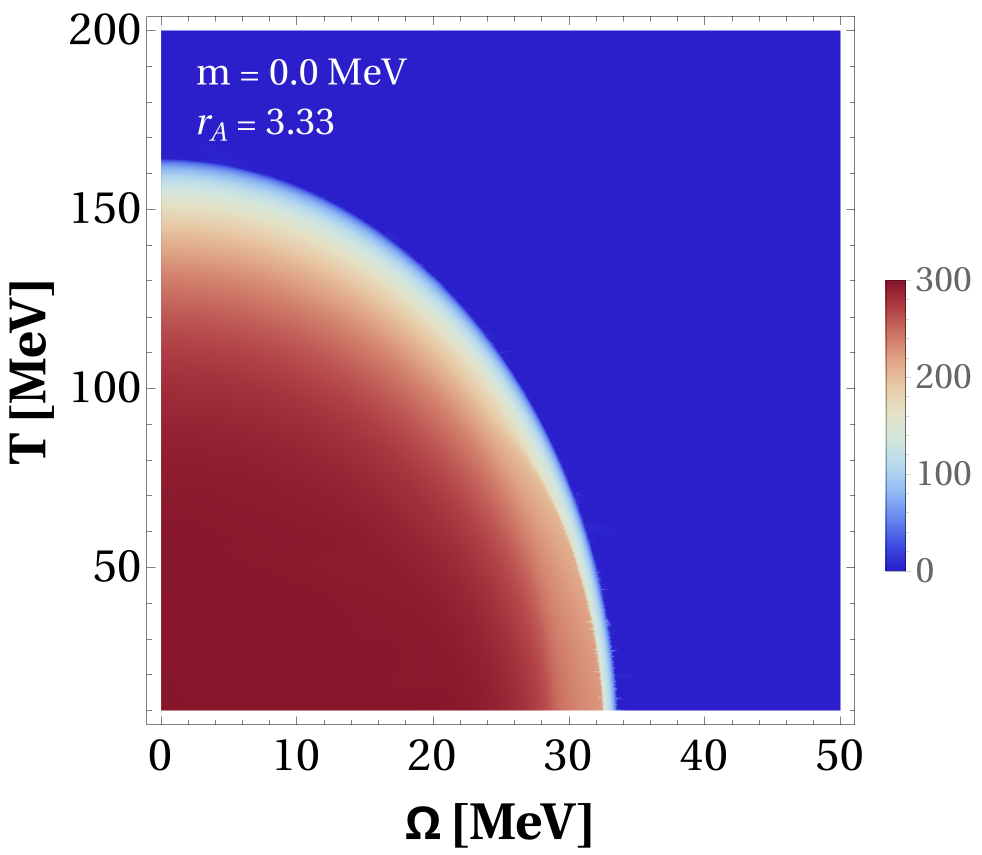}
\includegraphics[scale=0.3]{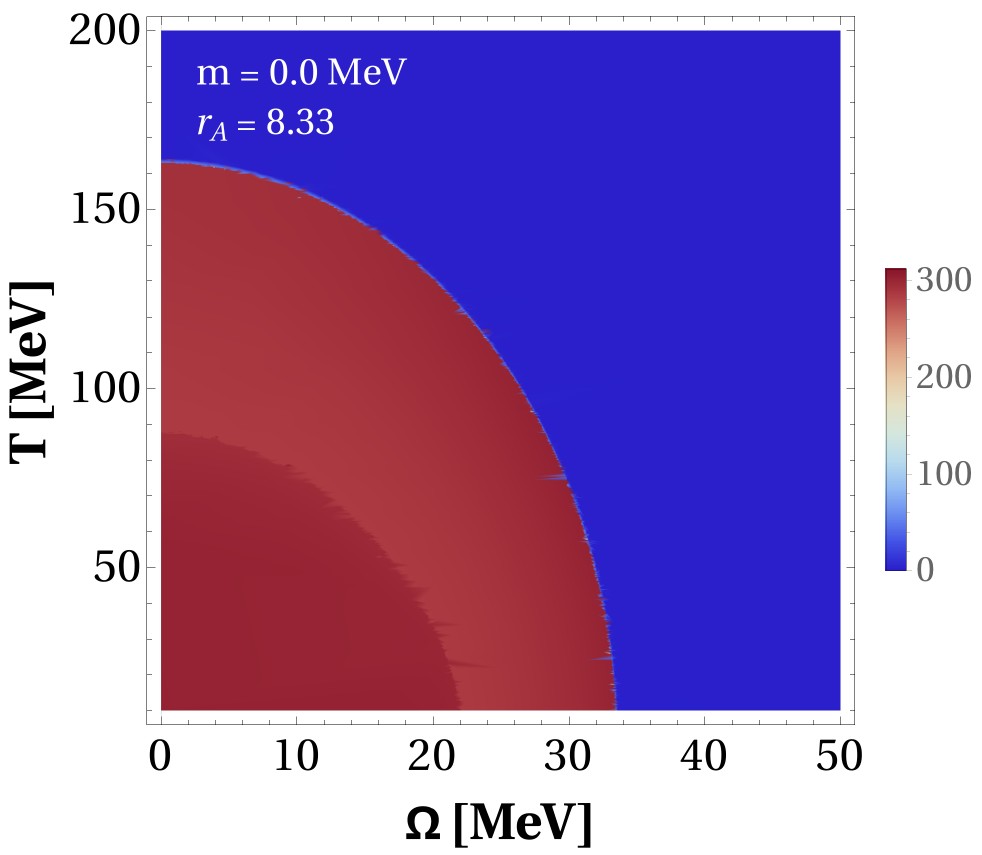}\\
\includegraphics[scale=0.3]{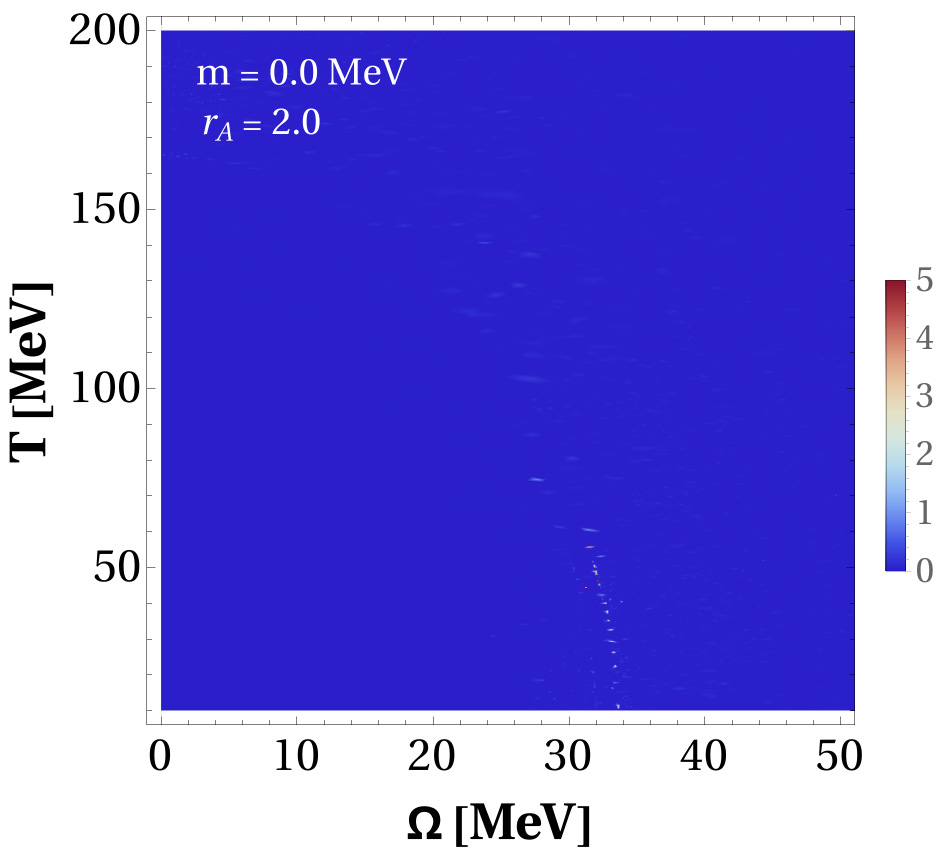}
\includegraphics[scale=0.3]{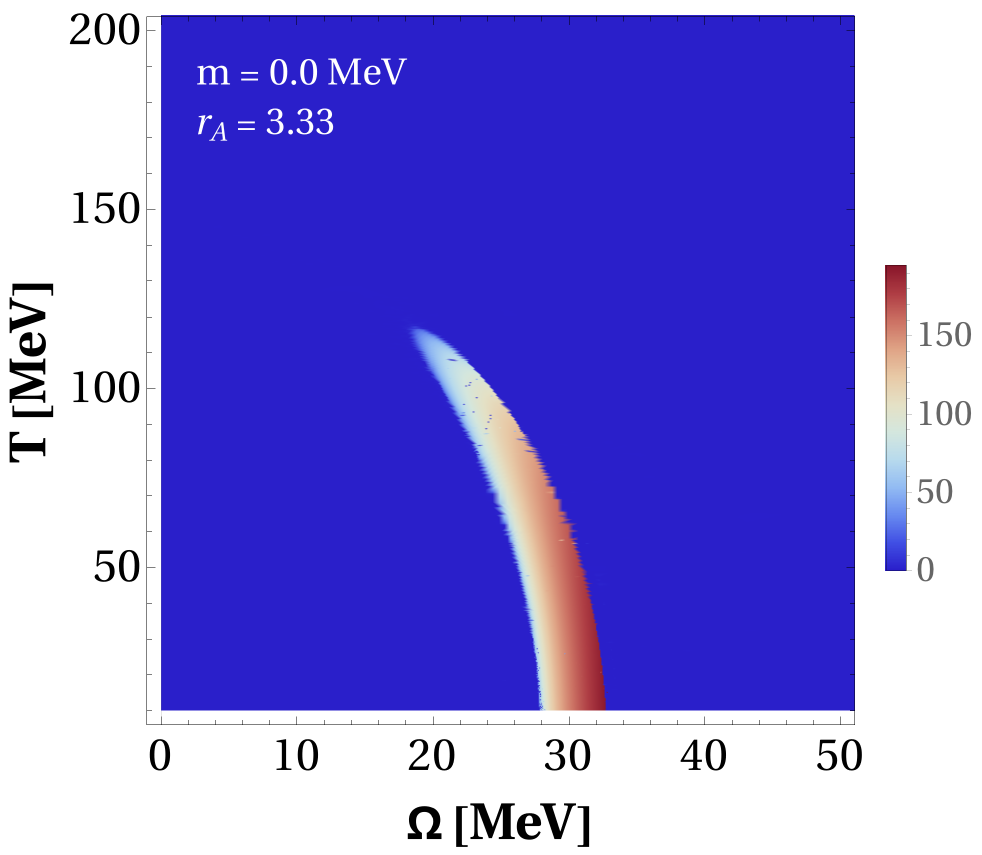}
\includegraphics[scale=0.3]{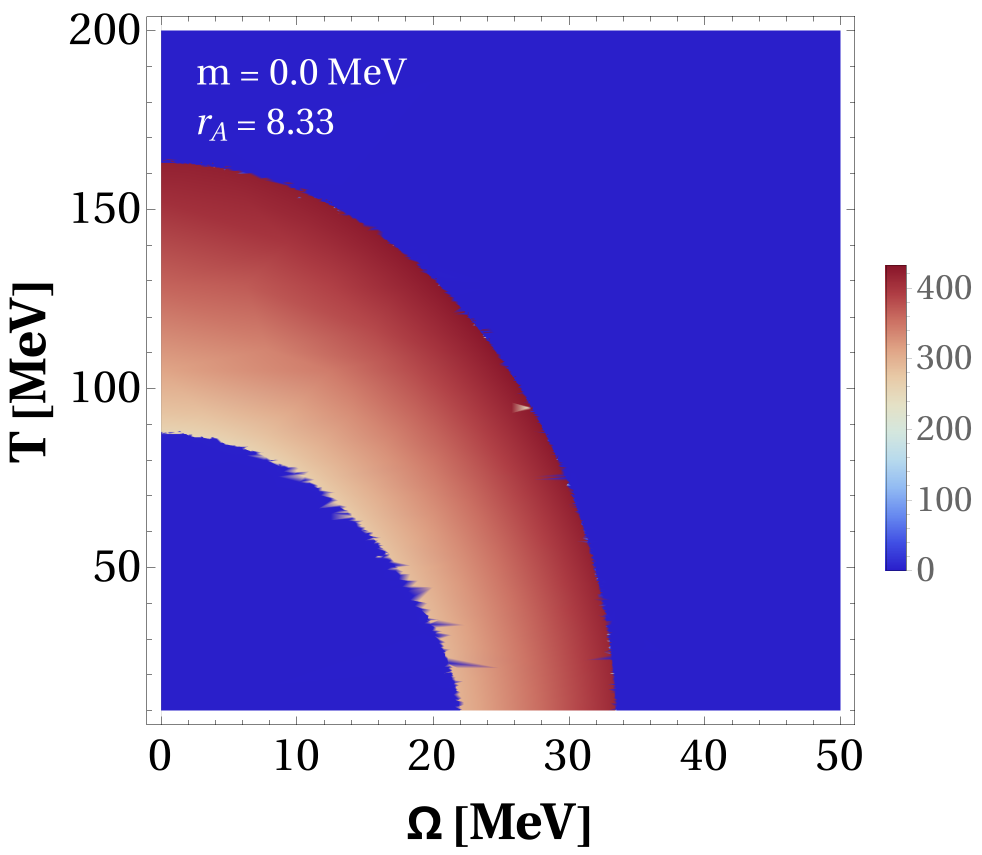}
\caption{ Top: Contour plots of the chiral condensate $\sigma$ in the $T-\Omega$ plane for $r_A = 2.0$ (left),  $r_A = 3.33$ (middle) and $r_A = 8.33$ (right). Bottom: The respective contour plots of the spin condensate $\mathbf{s}$ in the $T-\Omega$ plane.
 }\label{Fig:Phasediagram_1}
\end{figure*}

From \cref{eq:Potential_final}, it is evident that the medium-dependent part of the effective potential contributes negatively. 
This means that  increasing the magnitude of this term enhances the negative contribution and may lead to a shift of the minimum of the potential. 
Comparing case (a), \(\sigma = \mathbf{s} = 0\), with case (b), \(\sigma = 0\), \(|\mathbf{s}| = 200\, \text{MeV}\), we find by an explicit calculation that the negative medium-dependent term in the effective potential \(V_\text{eff}\) (cf.~\cref{eq:Potential_final}) does not change at all.
However, the positive contribution proportional to \(\mathbf{s}^2\) increases in case (b) due to the nonvanishing spin condensate \(|\mathbf{s}| = 200\, \text{MeV}\). 
Thus, a nonzero spin condensate  \(\mathbf{s}\) is never energetically favored, unless the chiral condensate \(\sigma\) is also nonvanishing, like in cases (c) and (d).
 To fully understand the interplay between the spin and chiral condensates, the minimum of the effective potential \(V_\text{eff}\) must be found self-consistently.\\

\textit{Phase diagram of $\sigma$ and $\mathbf{s}$ at finite $T$}: It is known that, in effective models, the global rotation of the system, corresponding to a nonvanishing macroscopic orbital angular momentum, induces a rotational polarization effect that tends to align microscopic angular momenta with the global rotation axis \cite{Jiang_2016, Chernodub:2016kxh, Wang:2018sur, Chen:2020ath,  Golubtsova:2022ldm, Chen:2022smf, Singha:2024tpo, Jiang:2023zzu, Sun:2023kuu, Chen:2023cjt, Zhao:2022uxc, Yadav:2022qcl, Braga:2022yfe, Mehr:2022tfq, Sadooghi:2021upd, Fujimoto:2021xix, Zhang_Boundary, Chen:2022mhf, Nunes:2024hzy}. 
In the case of quark and antiquark pairing, both their relative orbital angular momentum $L$ and spin $S$ are driven to align with the global angular momentum, rather than forming a scalar state with total angular momentum $J=0$. 
This mechanism generically suppresses scalar pairing states, such as the chiral condensate composed of quark-antiquark pairs with $L=1$, $S=1$, but $J=0$. 

In contrast, if a spin condensate forms corresponding to quark-antiquark pairs with total angular momentum $J= 1$, the rotational polarization effect enhances the stability of such a phase, as both spin and orbital components align with the global angular momentum. 
This has important implications for chiral symmetry restoration at high angular momentum: the spin condensate can ``absorb'' angular momentum that would otherwise disrupt the chiral condensate. 
Moreover, as discussed previously, in the absence of spin condensation, chiral symmetry is restored at sufficiently large angular momentum, at least for an infinite system. 
In the zero-temperature limit (\( T \to 0 \)), spin condensation does not occur when the chiral condensate \( \sigma \to 0 \), see discussion of case (b) above. 
Therefore, a competition arises between angular momentum, which promotes chiral-symmetry restoration, and spin condensation, which tends to break it. 
To explore this interplay quantitatively, we must solve the coupled gap equations self-consistently.
\begin{align} \label{eq:gap_equations}
    \frac{\partial V_{\mathrm{eff}}}{\partial \sigma} \overset{!}{=} 0\;, \quad \frac{\partial V_{\mathrm{eff}}}{\partial \mathbf{s}} \overset{!}{=} 0\; .
\end{align}

We numerically solve the gap equation with $N_c=3$ and $N_f=2$, and choose the parameter $G=6.002/\Lambda^2$, where $\Lambda=757.048$ MeV \cite{ITPDarmstadt,Klevansky1992}.
We fix these parameters to reproduce the constituent quark mass in vacuum, $\sigma=300$ MeV ($T=0$, $\Omega=0$ and $\mathbf{s}=0$), and the pion decay constant $f_\pi=88$ MeV in the chiral limit. 
We also define the parameter $r_A\equiv G_A/G$.

In \cref{Fig:Phasediagram_1}, the top row presents contour plots of the chiral condensate in the $T-\Omega$ plane for different parameters $r_A$ in the chiral limit, while the bottom row presents the respective contour plots of the spin condensate.
The left column shows the behavior of the chiral and spin condensates for $r_A = 2.0 $, the middle panel for $r_A = 3.33 $, and the right panel  for $r_A = 8.33 $. 
For a small $r_A=2.0$, the chiral condensate gradually decreases with increasing temperature \(T\) and increasing $\Omega$, and vanishes in a second-order phase transition.
The critical temperature at $\Omega = 0$ is  \(T_c \simeq 165\) MeV, while the critical angular velocity at $T= 0$ is \(\Omega_c \simeq 33\) MeV. 
At the same time, we do not find regions with a finite spin condensate.

This behavior changes when $r_A$ is increased. 
For $r_A=3.33$ (middle panels of \cref{Fig:Phasediagram_1}), we observe the emergence of a region at low temperatures and intermediate angular velocities where the spin condensate assumes a finite value.
This region is bounded towards smaller values of $\Omega$ by a second-order phase transition and towards larger values by a first-order phase transition, see detailed discussion of \cref{Fig:cut_Plot} below.
For even larger values of $r_A$ the region of a nonvanishing spin condensate extends all the way to the temperature axis, cf.~lower right panel of \cref{Fig:Phasediagram_1}.
Both boundaries exhibit first-order phase transitions.
The chiral condensate (upper right panel of \cref{Fig:Phasediagram_1}) is now substantially influenced by the spin condensate, cf.~detailed discussion of \cref{Fig:cut_Plot} below.

\begin{figure*}
\includegraphics[scale=0.3]{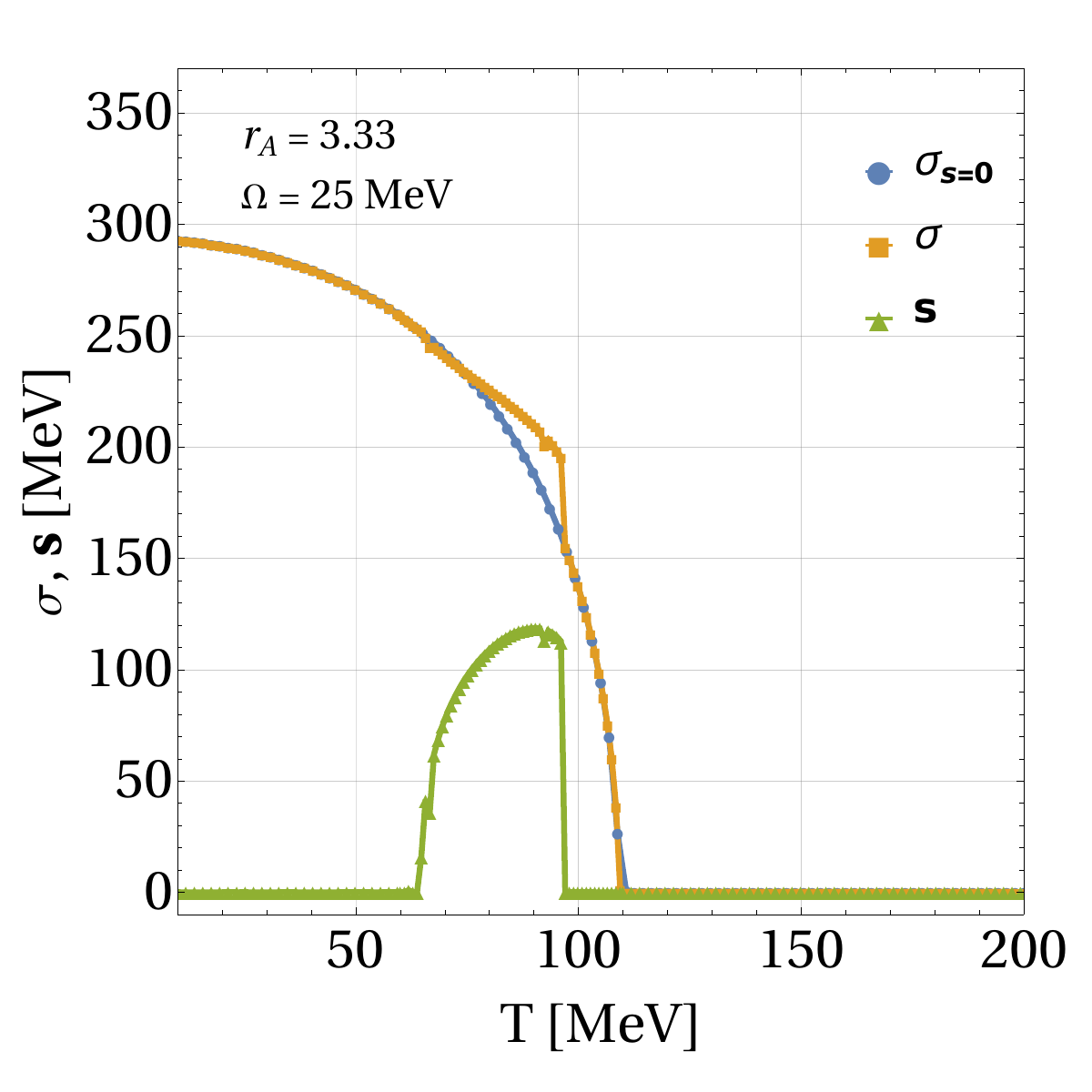}
\includegraphics[scale=0.3]{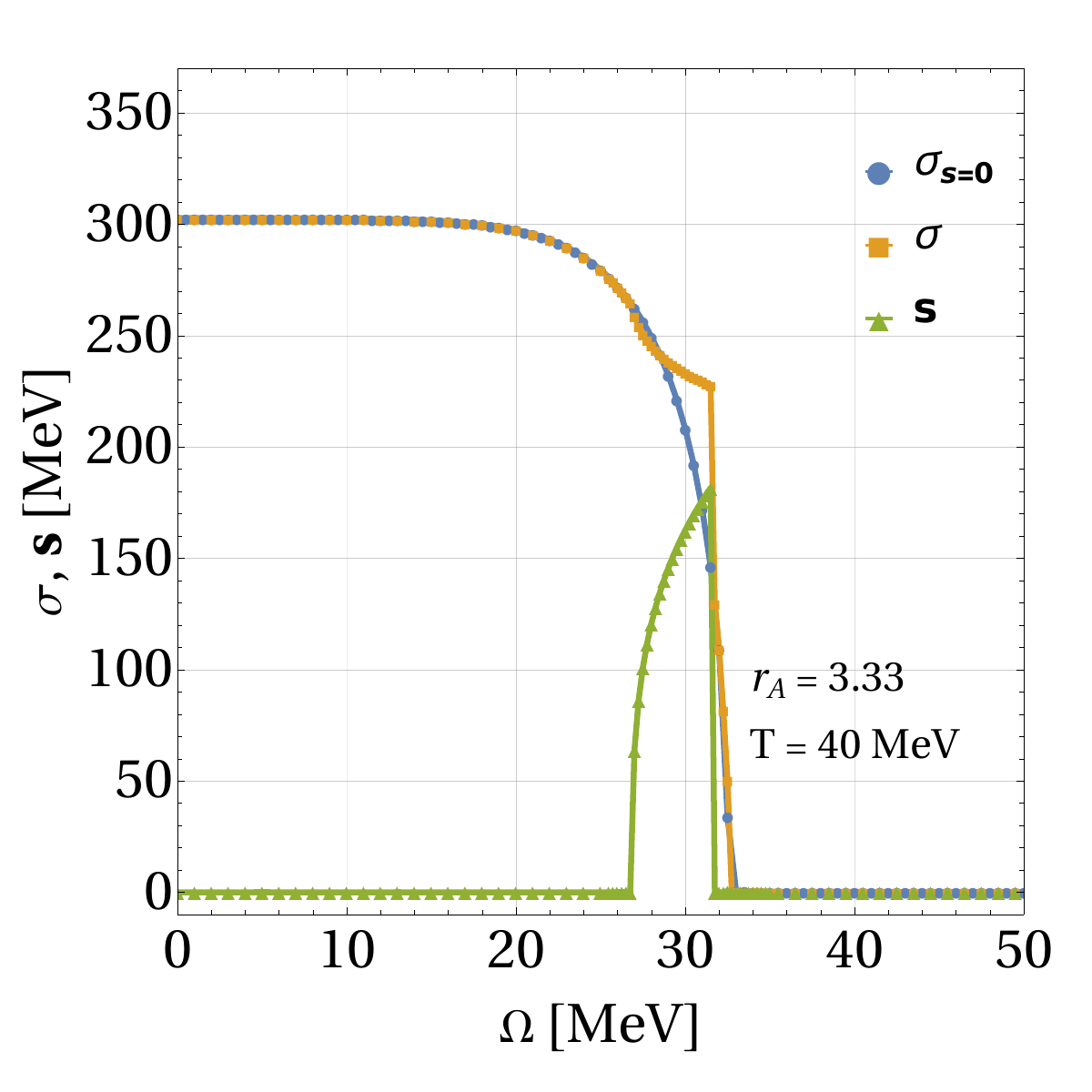}\\
\includegraphics[scale=0.3]{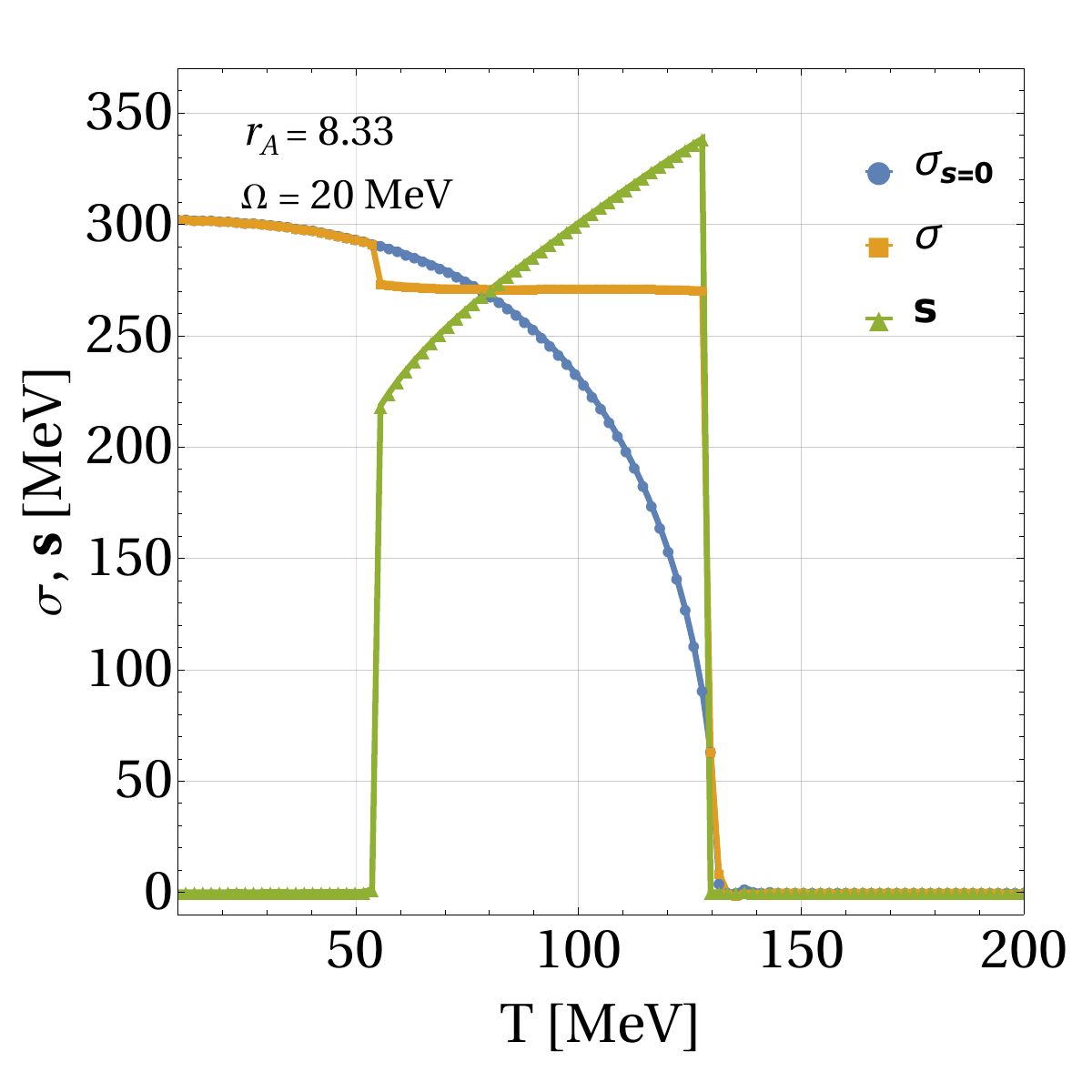}
\includegraphics[scale=0.3]{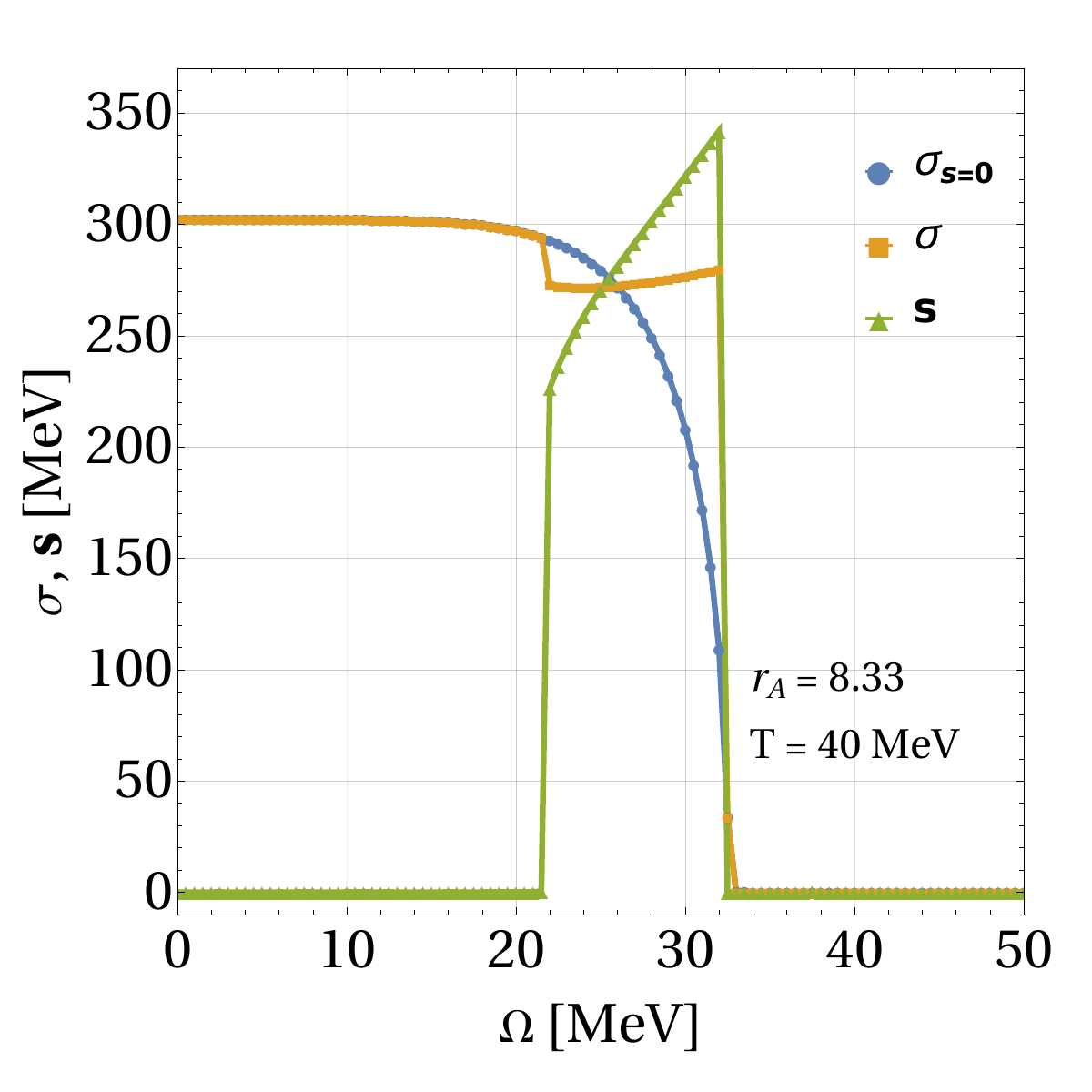}
\caption{The upper row correspond to results for $r_A = 3.33$ and the lower row to $r_A = 8.33$, respectively. 
Left panels: temperature dependence of the condensates at fixed angular velocity $\Omega = 25~\mathrm{MeV}$ (top) and $\Omega = 20~\mathrm{MeV}$ (bottom). 
Right panels: angular-velocity dependence of the condensates at fixed temperature $T = 40~\mathrm{MeV}$. 
The blue curve $\sigma_{\mathbf{s}=0}$ denotes the chiral condensate in the absence of a spin condensate, 
the orange curve $\sigma$ represents the chiral condensate, 
and the green curve shows the spin condensate $|\mathbf{s}|$, obtained from a solution of the coupled gap equations.
}\label{Fig:cut_Plot}
\end{figure*}
In \cref{Fig:cut_Plot} we show the \(T\)-dependence (left column) and the \(\Omega\)-dependence (right column) of the condensates. 
The top row corresponds to $r_A=3.33$, while the bottom row is for $r_A=8.33$, respectively.
For the temperature dependence of the condensates (left column) we fix $\Omega=25$~MeV for $r_A=3.33$ and $\Omega=20$~MeV for $r_A=8.33$, while for the angular-velocity dependence we choose $T=40$~MeV in both cases. 
As already discussed in the context of \cref{Fig:Phasediagram_1}, without spin condensation the chiral condensate (blue line) undergoes a second-order phase transition in both $T$ and $\Omega$ directions. 
For $r_A=3.33$, the spin condensate appears in a second-order phase transition at smaller values of $T$ and $\Omega$, while for $r_A = 8.33$, this transition is of first order.
In both cases, the spin condensate vanishes in a first-order phase transition at larger values of $T$ and $\Omega$. 
The values of $T$ and $\Omega$ where this occurs are smaller than the values $T_c$ and $\Omega_c$ where the chiral condensate vanishes in a second-order phase transition in the case without spin condensation.
Wherever the spin condensate is zero, the chiral condensate (orange line) is identical to its value without spin condensation (blue line).
However, when the spin condensate is nonvanishing, it influences the behavior of the chiral condensate.
For instance, when the spin condensate appears or vanishes in a first-order transition, there is also a discontinuity in the chiral condensate.
In both cases, there is a region of ``rotational suppression'' at smaller values of $T$ and $\Omega$, where the chiral condensate is reduced, and a region of ``rotational enhancement,'' where it is increased as compared to the case without spin condensation.
This ``(inverse) rotational catalysis'' bears a conceptual analogy to (inverse) magnetic catalysis, where an external magnetic field strengthens (diminishes) the chiral condensate \cite{Shovkovy2007, Gusynin_1995, Andersen2014, Bali2012, Miransky2015}.
 \begin{figure*}[!t]
 \includegraphics[scale=0.3]{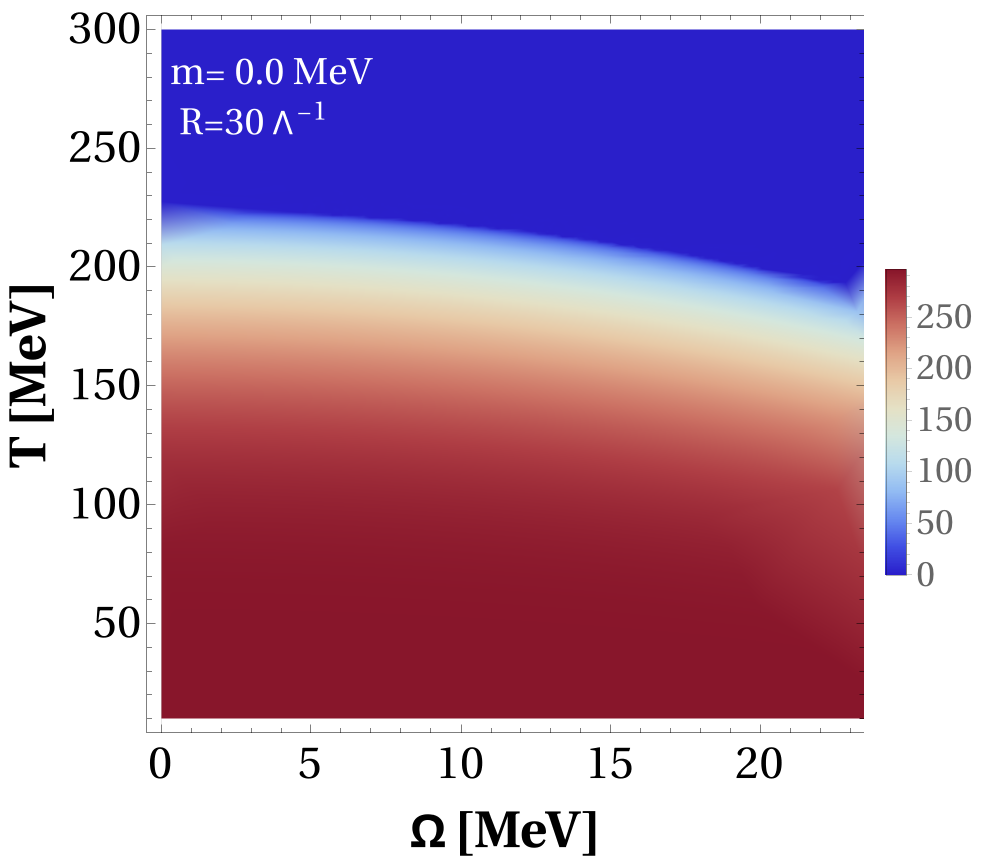}
\includegraphics[scale=0.3]{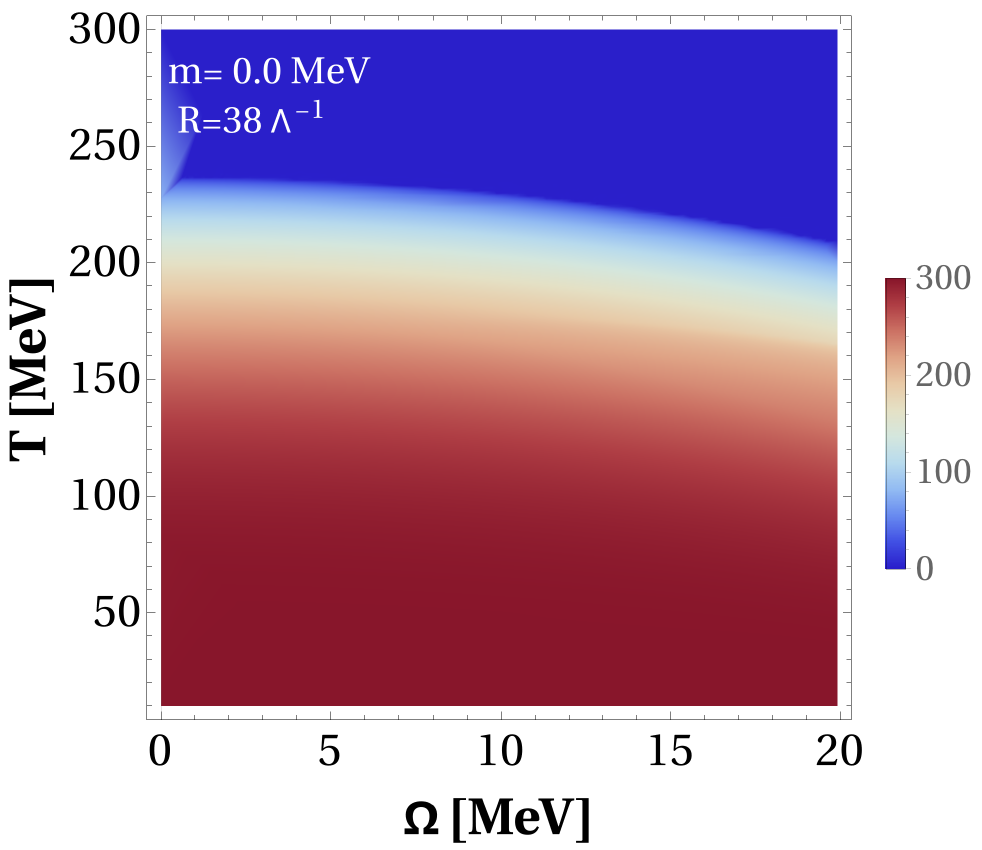}
\includegraphics[scale=0.3]{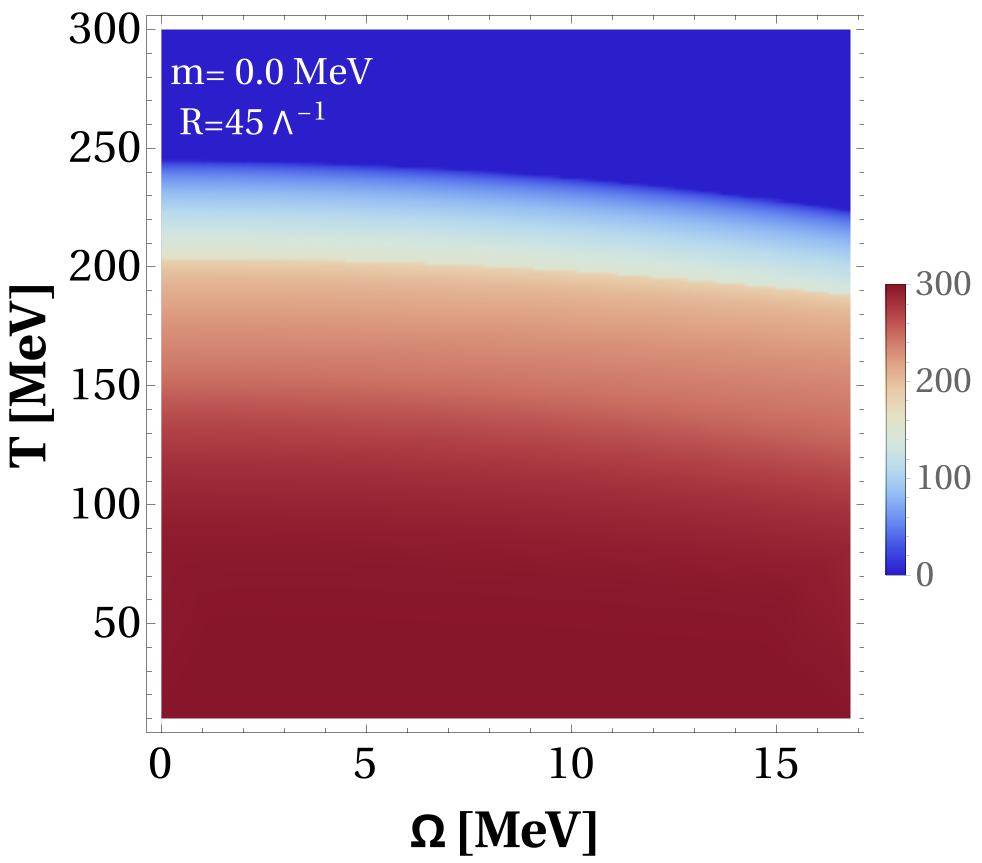}\\
\includegraphics[scale=0.3]{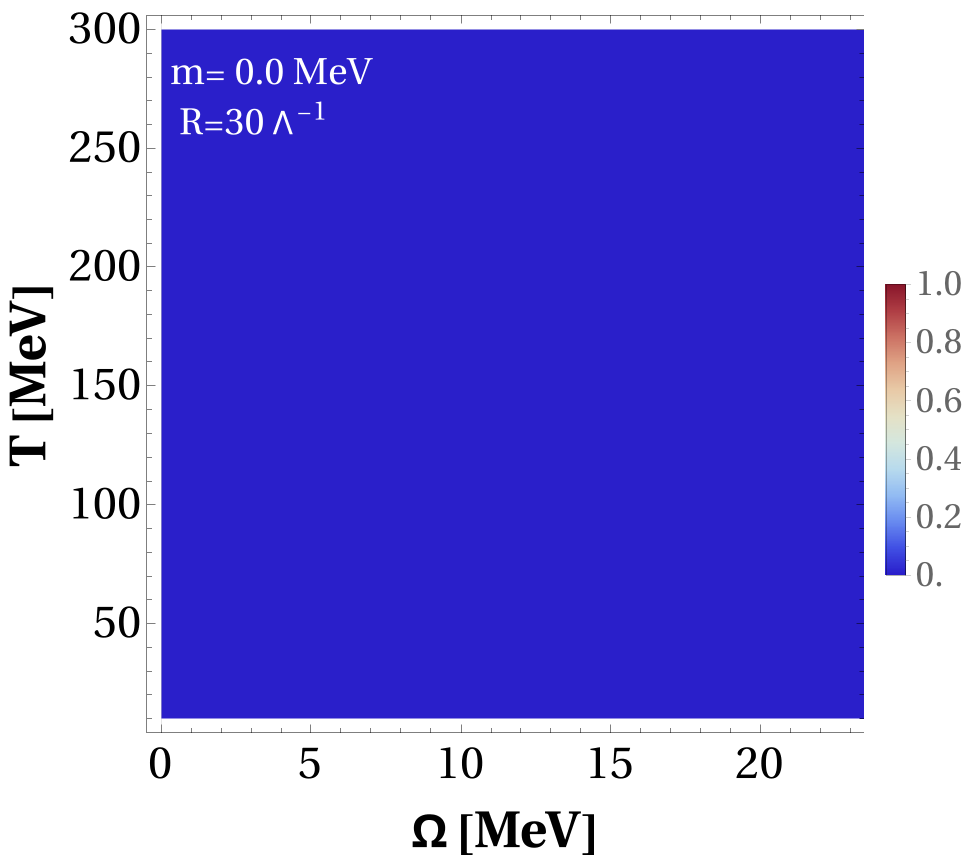}
\includegraphics[scale=0.3]{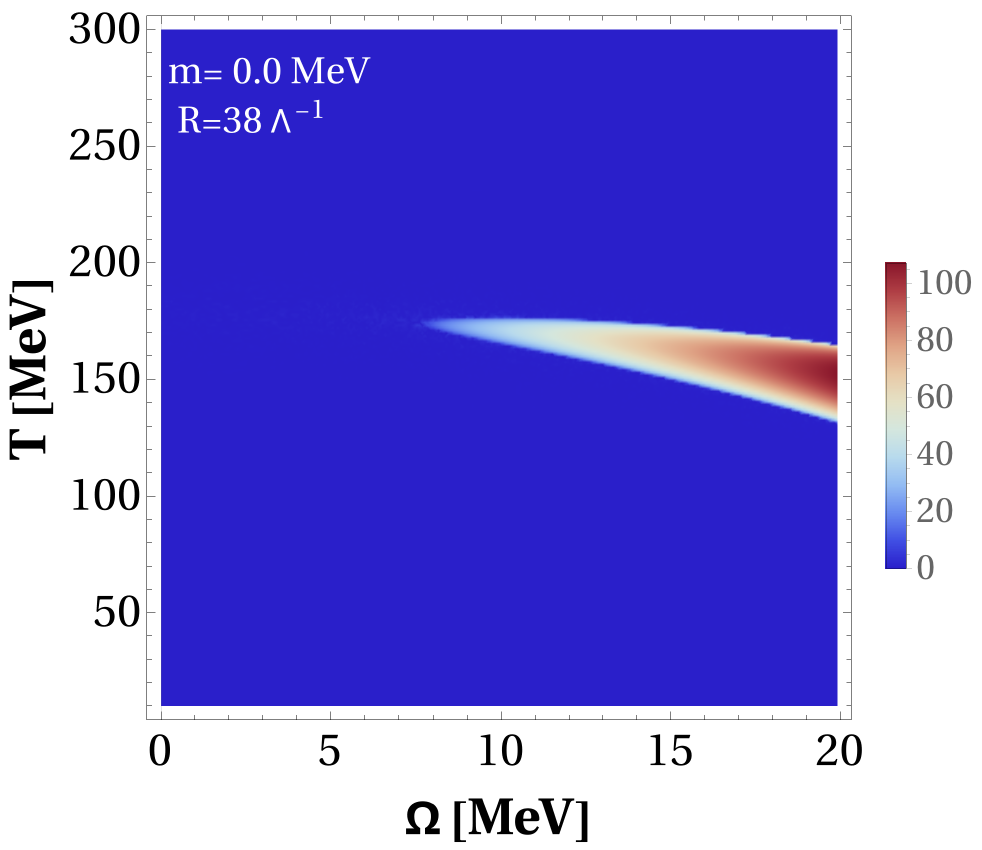}
\includegraphics[scale=0.3]{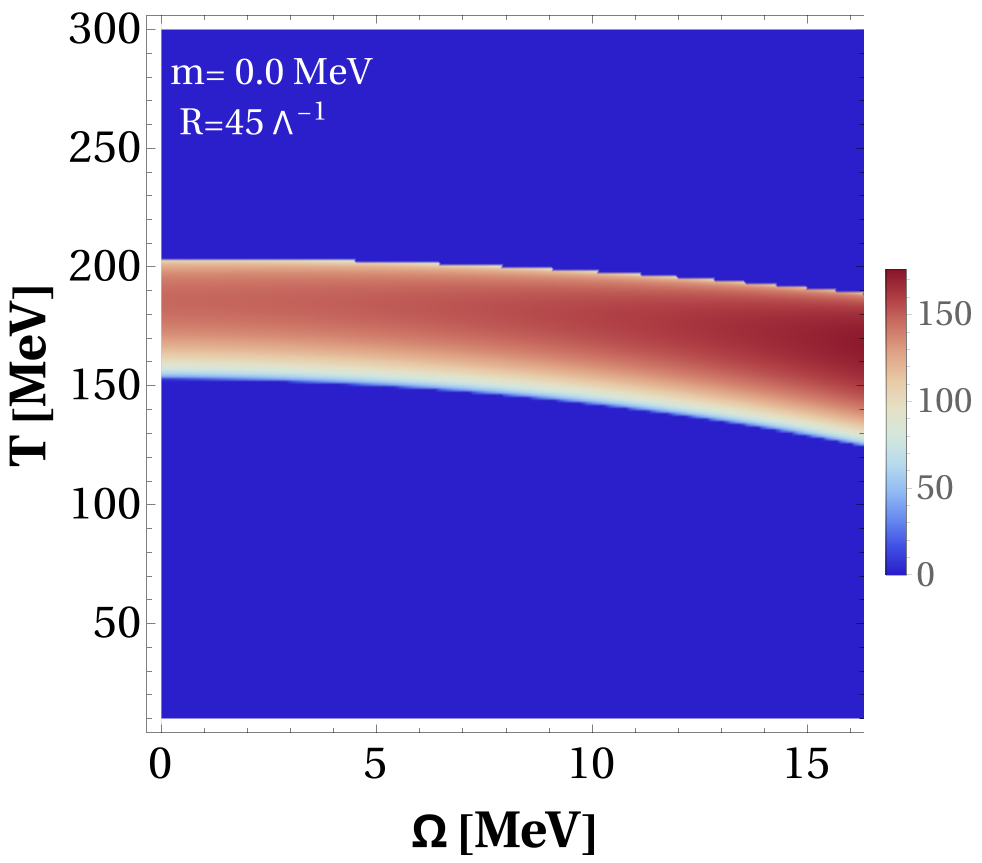}
\caption{Top: Contour plots of the chiral condensate $\sigma$ in the $T-\Omega$ plane for $r_A =4.76$ and different system sizes $R=30\Lambda^{-1}$ (left), $R=38\Lambda^{-1}$ (middle)  and $R=45\Lambda^{-1}$ (right). Bottom: The respective contour plots of the spin condensate $\mathbf{s}$ in the $T-\Omega$ plane.
 }\label{Fig:Phasediagram_BC}
\end{figure*}

\textit{Effect of boundary conditions}: 
In a system that rotates rigidly with angular velocity $\Omega$, fluid elements at radial distances $r> \Omega^{-1}$ assume  superluminal velocities $v_r>1$, violating causality. 
Therefore it is necessary to impose boundary conditions to ensure that $v_r \leq 1$.
In our case we assume that the radial extent of the system is limited, $r \leq R = \Omega^{-1}$\cite{Mameda_Boundary, Ambru_Boundary, Zhang_Boundary}. 
Imposing these boundary conditions leads to a discretization of the radial momentum $p_\perp$ :
\begin{align}
    p_\perp \rightarrow  p_{l,k}=\begin{cases}
    \frac{\xi_{l,k}}{R}\quad ,\,l = 0,1,2,\dots\,, \\
    \frac{\xi_{-l-1,k}}{R} \quad ,\, l = -1,-2, \dots\, ,
\end{cases}
\end{align}
where $\xi_{l,k}$ is the $k$-th positive root of the cylindrical Bessel function $J_l(x)$.
This has the consequence that the integral over $p_\perp$ in \cref{eq:Potential_final} is replaced by a sum over $k$, resulting in:

\begin{widetext}
\begin{align}\label{eq:Potential_boundary_final}
    V_{\text{eff}}&=\frac{(\sigma-m)^2}{2G}+\frac{\mathbf{s}^2}{2G_A}-\frac{N_f N_c}{(2\pi)^3}\sum_{k=1}^\infty\sum_{l=0}^\infty\sum_{e=\pm} \sum_{s=\pm}\int dp_z  \frac{2}{N_{e\cdot l,k}^2}\bigg[ T\,\ln(1+e^{-(E_s-e\Omega l)/T}) 
    +E_s\bigg]\,,
\end{align}
with the normalization factor
\begin{align}
    N_{e\cdot l,k}=R \cdot \begin{cases}
    |J_{l+1}(\xi_{l,k})| \quad , \,e\cdot l = 0,1,2,\dots \, , \\
    |J_{-l}(\xi_{-l-1,k})|  \quad ,\, e\cdot l = -1,-2, \dots \, .
\end{cases}
\end{align}
\end{widetext}

\begin{figure*}[ht]
\color{black}
\includegraphics[scale=0.3]{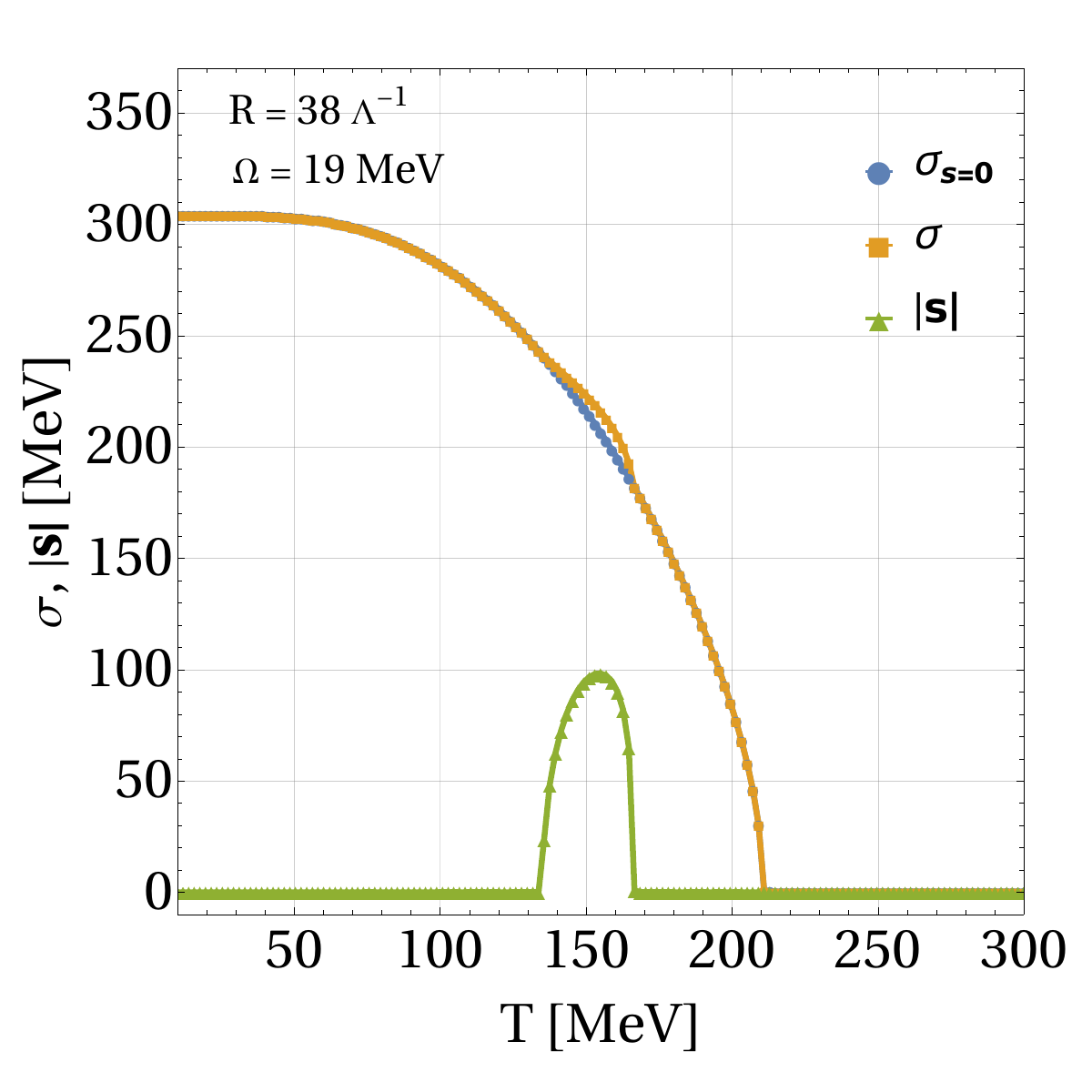}
\includegraphics[scale=0.3]{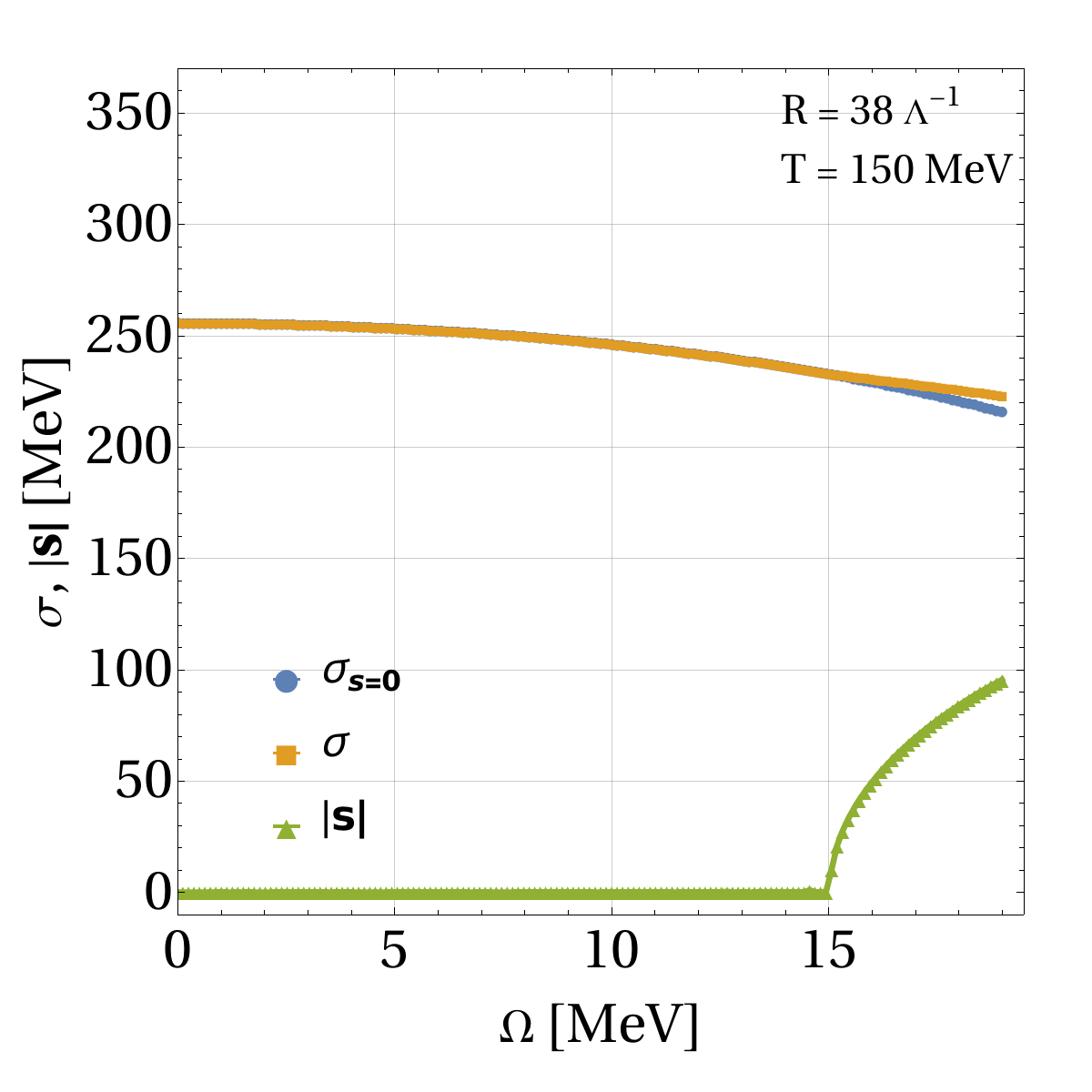}\\
\includegraphics[scale=0.3]{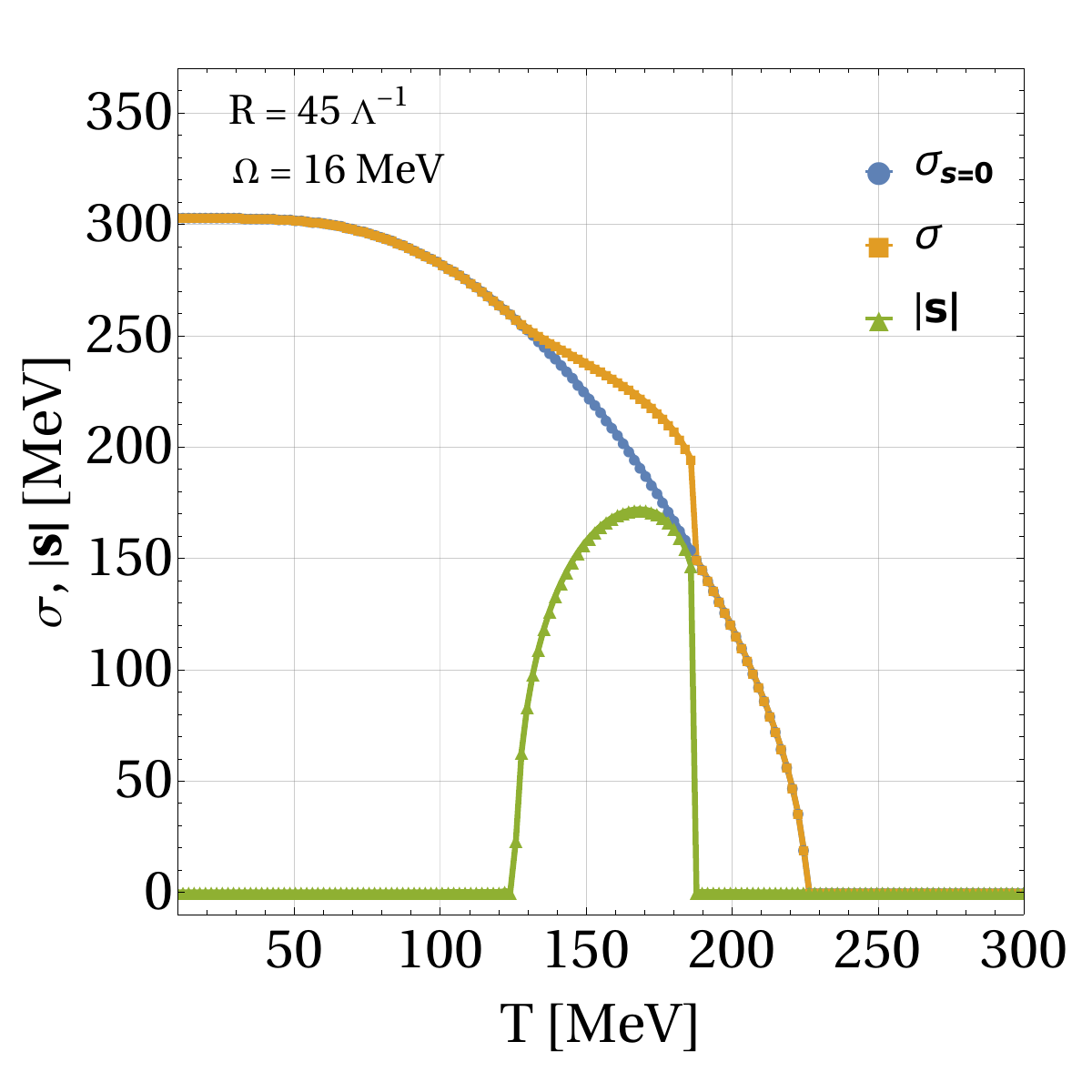}
\includegraphics[scale=0.3]{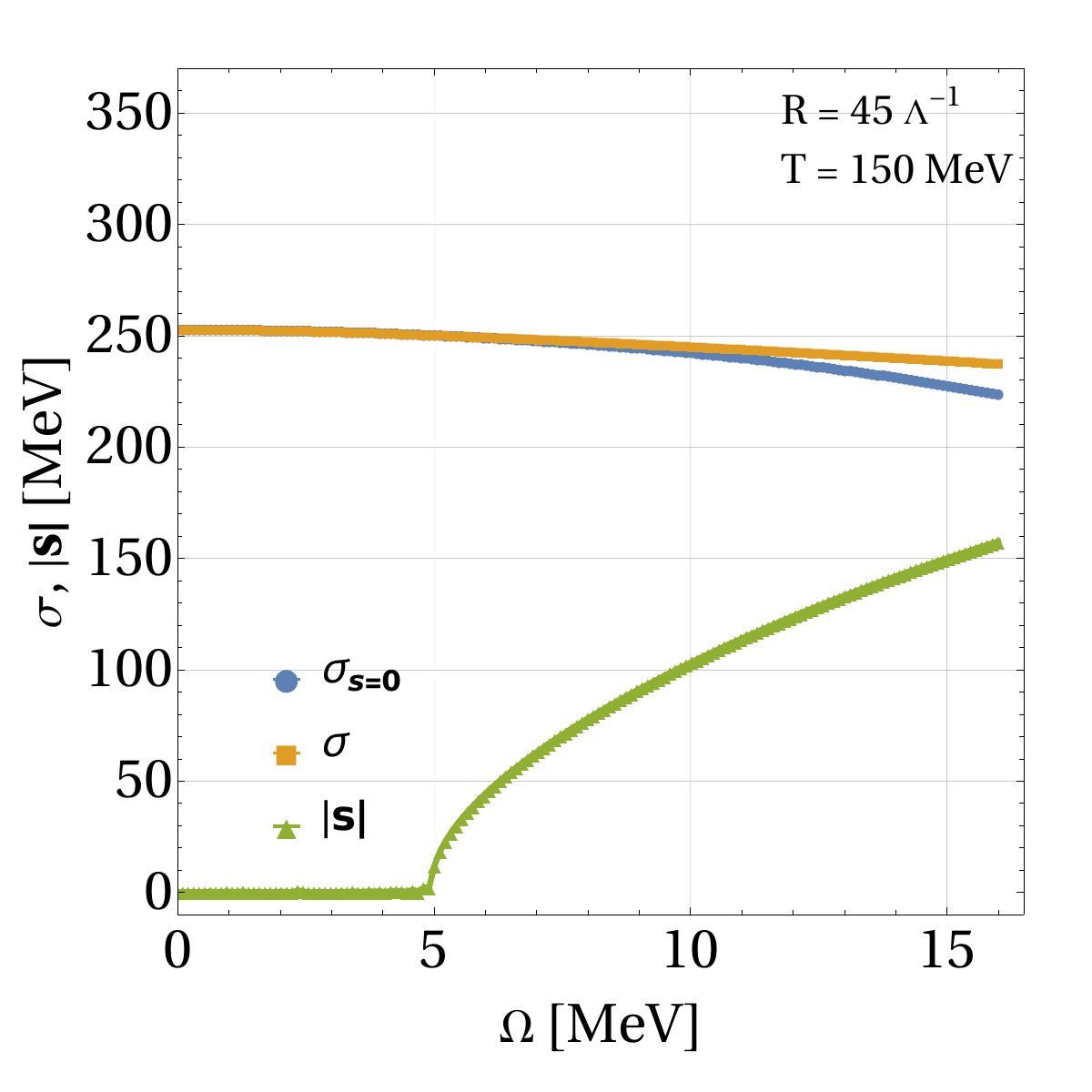}
\caption{The upper row correspond to results for radius $R = 38\Lambda^{-1}$ and the lower row to $R = 45\Lambda^{-1}$, respectively, for $r_A =4.76$. 
Left panels: temperature dependence of the condensates at fixed angular velocity $\Omega = 19~\mathrm{MeV}$ (top) and $\Omega = 16~\mathrm{MeV}$ (bottom). 
Right panels: angular-velocity dependence of the condensates at fixed temperature $T = 150~\mathrm{MeV}$. 
The blue curve $\sigma_{\mathbf{s}=0}$ denotes the chiral condensate in the absence of a spin condensate, 
the orange curve $\sigma$ represents the chiral condensate, 
and the green curve shows the spin condensate $|\mathbf{s}|$, obtained from a solution of the coupled gap equations.
}\label{Fig:Phasediagram_BC1}
\end{figure*}

\textit{Phase diagram of $\sigma$ and $\mathbf{s}$ at finite $T$ with boundary conditions}: It is well known that imposing boundary conditions modifies the phase structure dramatically, especially at lower temperatures. 
For instance, in Ref.~\cite{Mameda_Boundary} it has been shown that at zero temperature, there is no chiral symmetry restoration at all  when increasing the angular velocity. 
In the following, we study the influence of boundary conditions on both the chiral as well as the spin condensate.

In \cref{Fig:Phasediagram_BC}, similar to the previous section, the top row shows contour plots of the chiral condensate in the $T-\Omega$ plane for different radii $R$ in the chiral limit, while the bottom row displays the corresponding contour plots of the spin condensate. 
The left column presents the results for $R = 30\,\Lambda^{-1}$, the middle column for $R = 38\,\Lambda^{-1}$, and the right column for $R = 45\,\Lambda^{-1}$, corresponding to approximately $7.8$ fm, $10.0$ fm, and $11.7$ fm, respectively. 
These radii correspond to typical system sizes encountered in heavy-ion collisions and are therefore of physical relevance. 
In this comparison of systems with different sizes, we fix the ratio $r_A = 4.76$.
Our first, and most important, observation is that at low temperatures chiral symmetry is never restored for any value of $\Omega$ up to the maximum value allowed by causality.
This confirms the results of Ref.~\cite{Mameda_Boundary}.

For smaller radii, the chiral condensate gradually decreases with increasing temperature $T$ and vanishes via a second-order phase transition. 
As $\Omega$ increases, the chiral condensate remains nearly constant for temperatures below $T \simeq 150\,\text{MeV}$ over the full range of $\Omega$. 
Only close to the phase boundary does the chiral condensate $\sigma$ exhibit a noticeable dependence on $\Omega$. 
This qualitative behavior is observed for all radii $R$ considered. 
The main quantitative difference between the phase diagrams for different radii lies in the critical temperature $T_c(\Omega)$, which increases with $R$. 
For $\Omega = 0$, we find $T_c \simeq 220\,\text{MeV}$ (left), $T_c \simeq 230\,\text{MeV}$ (middle), and $T_c \simeq 240\,\text{MeV}$ (right).
This effect is obviously absent in an infinite system.

For the smallest radius, $R = 30\,\Lambda^{-1}$, no spin condensate is observed. 
This changes for larger radii. 
For $R = 38\,\Lambda^{-1}$, spin condensation sets in within a small region for angular velocities $\Omega \gtrsim 8\,\text{MeV}$ and temperatures $130\,\text{MeV} \leq T \leq 180\,\text{MeV}$. 
In this region, the chiral condensate $\sigma$ remains nearly constant and no longer exhibits a typical second-order phase transition. 
Instead, it undergoes a discontinuous jump to a different value at the point where the spin condensate $\mathbf{s}$ vanishes, reminiscent of the behavior found in the case of an infinite system without boundary conditions.

For the largest radius, $R = 45\,\Lambda^{-1}$, the spin condensate $\mathbf{s}$ exists over the entire range of angular velocities $\Omega$ considered. 
It appears in the temperature interval $120\,\text{MeV} \lesssim T \lesssim 210\,\text{MeV}$. 
As a function of temperature, the spin condensate sets in via a second-order phase transition and disappears either through a steep second-order transition at small $\Omega$ or through a first-order transition at large $\Omega$. 
Once again, in the region where spin condensation occurs, the chiral condensate remains nearly constant as a function of temperature and undergoes a discontinuous jump to a new value at the point where the spin condensate $\mathbf{s}$ vanishes.

In \cref{Fig:Phasediagram_BC1}, we show the temperature dependence (left column) and the angular-velocity dependence (right column) of the condensates. 
The top row corresponds to the radius $R = 38\,\Lambda^{-1}$, while the bottom row corresponds to $R = 45\,\Lambda^{-1}$. 
For the temperature dependence (left column), we fix the angular velocity to $\Omega = 19\,\text{MeV}$ for $R = 38\,\Lambda^{-1}$ and to $\Omega = 16\,\text{MeV}$ for $R = 45\,\Lambda^{-1}$.
For the angular-velocity dependence (right column), we choose $T = 150\,\text{MeV}$ in both cases.

As already discussed in the context of \cref{Fig:Phasediagram_BC}, in the absence of spin condensation the chiral condensate (blue line) undergoes a second-order phase transition as a function of temperature, while it changes only gradually as a function of $\Omega$ up to the maximum allowed angular velocity. 
For both $R = 38\,\Lambda^{-1}$ and $R = 45\,\Lambda^{-1}$, the spin condensate appears via a second-order phase transition and disappears via a first-order phase transition as a function of temperature, whereas it increases smoothly with $\Omega$ over the considered range.

We find that the chiral condensate $\sigma$ obtained when spin condensation is allowed coincides with the chiral condensate obtained without spin condensation in the region where no spin condensate is present (blue line), but it systematically assumes larger values in the presence of a nonzero spin condensate $\mathbf{s}$. 
This again demonstrates the phenomenon of ``rotational catalysis''. In contrast to the case without boundary conditions, however, we do not observe any region of inverse rotational catalysis. 
Moreover, the first-order transition of the spin condensate $\mathbf{s}$ induces a discontinuous jump in the chiral condensate $\sigma$.\\

\textit{Summary and Discussion}: 
In this work, we investigated the thermodynamic properties and phase behavior of spin degrees of freedom within the Nambu--Jona-Lasinio (NJL) model under rotation. 
We introduced an axial-vector interaction term into the NJL model to analyze the emergence of spin polarization and therefore the emergence of magnetized matter. 
Utilizing the metric of a rigidly rotating cylinder, we employed the mean-field approximation to calculate the effective potential, focusing on the coupled dynamics of chiral and spin condensates in an infinite as well as a finite system.

The results of our study reveal several notable features. 
Firstly, the inclusion of rotation induces non-trivial effects on quark matter, such as spin alignment analogous to the Barnett effect. 
This alignment, driven by rotational dynamics, manifests as anisotropies in momentum space and deformations of the Fermi surface. 
Secondly, the interplay between the chiral and spin condensates under rotation results in a region of nonvanishing spin condensate in the plane of temperature and angular velocity  for the infinite and the finite system.

For the  infinite system, depending on the value of the axial coupling constant, this region is bounded by second- and/or first-order phase transitions.
At the first-order phase transition lines, the chiral condensate also exhibits a discontinuity.
Within this region, we observe ``(inverse) rotational catalysis'' of the chiral condensate.

In contrast, in the finite system we never observe ``inverse rotational catalysis'' but always ``rotational catalysis''. 
Another difference is that the spin condensate always appears in a second-order, but never in a first-order phase transition. Similar to the infinite system the spin condensate vanishes either in a second- or first-order phase transition and the chiral condensate exhibits a discontinuity at the first-order phase transition of the spin condensate.

For the future, several open questions remain to be answered. 
Our study relied on the mean-field approximation, neglecting fluctuations and higher-order interactions that may play significant roles, especially near critical points. 
Future work could employ functional renormalization group methods to account for these effects and provide a more comprehensive description of the phase structure \cite{Schaefer_2008}. 
Furthermore, the inclusion of external magnetic fields and their interplay with rotational dynamics presents a promising direction for further exploration, potentially uncovering novel magnetohydrodynamics phenomena in quark matter \cite{PhysRevD.93.104052, Sadooghi_2021}. 
Moreover, the present work neglects the dynamics of gluons. However, in Refs.~\cite{Chernodub:2022veq, Braguta:2021jgn}, where gluonic degrees of freedom are included, the chiral transition temperature increases (rather than decreases) with increasing rotation. It is therefore expected that incorporating gluons could significantly modify the phase structure.

Finally, the emergence of a spin (axial) condensate entails the spontaneous breaking of rotational, boost, and parity symmetries along with chiral symmetry. 
According to Goldstone’s theorem, such symmetry breaking is expected to give rise to corresponding Nambu–Goldstone modes. 
These collective excitations -- analogous to magnons in condensed-matter systems where spin rotational symmetry is broken -- may manifest as long-lived hydrodynamic modes in the strongly interacting medium \cite{Watanabe_2011,Beekman_2019}. 
Their properties, dispersion relations, and role in angular-momentum transport deserve systematic exploration within an extended framework of relativistic (spin) hydrodynamics.

In summary, this work represents a step forward in understanding the intricate dynamics of spin-polarized quark matter under rotation. 
By establishing a connection between fundamental spin interactions and macroscopic phase behavior, we contribute to the broader effort of unraveling the rich physics of  QCD in extreme environments.\\

\textit{Acknowledgments}: We thank Matteo Buzzegoli, Chowdhury Aminul Islam,  Andrea Palermo, Johannes Poeplau, and Masoud Shokri for useful discussions. 
We also thank Kazuya Mameda for bringing Ref.~\cite{Mameda_Boundary} to our attention, which prompted us to add the  results for a finite system with appropriate boundary conditions.
This work is supported by the \textit{Deutsche Forschungsgemeinschaft} (DFG, German Research Foundation) through CRC-TR 211 \textit{"Strong-interaction matter under extreme conditions"} - project number 315477589 - TRR 211 and by the State of Hesse within Research Cluster ELEMENTS (Project No.~500/10.006). 

\textit{Data availability}: The data used to generate the figures in this work can be found in \cite{Data}.

\section*{Supplementary Material}

\subsection*{Effective potential in the mean-field approximation}
In the supplementary material, we provide details of the derivation of the effective potential for an infinite system.
As explained in the main text, the action of the NJL model after a Hubbard-Stratonovich transformation and in mean-field approximation reads
\begin{align}
    S\left[ \sigma, s^\mu \right] &=\int d^4x \sqrt{g}\, \bar{\psi}(x) \, \left( i \gamma^a D_a - \sigma - \gamma^3 \gamma^5 |\mathbf{s}| \right)\, \psi(x)\nonumber\\
    & - \frac{V_4}{2G} (\sigma - m) ^2 - \frac{V_4}{2 G_A} \mathbf{s}^2\,  ,
\end{align}
with $g = \det(e_a^\mu)$ and $V_4$ is the volume of the system.
Without loss of generality we choose the components of the spin condensate $s^\mu$ orthogonal to the rotation axis $z$ to be zero.
Performing a Wick rotation $t = -i \tau$ yields the Euclidean action:
\begin{align}
    S_E\left[ \sigma, s^\mu \right] &=\int d^3x\,\int_0 ^\beta d\tau  \sqrt{g_E}\, \bar{\psi}(x)\, \big( i \gamma_E^a D_{E,a} - \sigma   \nonumber\\
    &- \gamma_E^3 \gamma_E^5 |\mathbf{s}| \big)\, \psi(x) - \frac{V_4}{2G} (\sigma - m) ^2 - \frac{V_4}{2 G_A} \mathbf{s}^2\,  \, .
\end{align}
From this the effective potential follows as:
\begin{align} \label{eq:Veff_1}
    V_{\text{eff}}\left(\sigma, |\mathbf{s}|\right) &= \frac{1}{2G} \left(\sigma - m \right) ^2 + \frac{1}{2 G_A} \mathbf{s}^2 \nonumber\\
    &+ \frac{i}{V} \Tr{\ln{(i \gamma_E^a D_{E,a}-\sigma - \gamma_E^3\gamma_E^5|\mathbf{s}|)}} \;,
\end{align}
where the $\Tr$ acts in color, flavor, Dirac, and coordinate space. 
Now we use the Ritus method \cite{RITUS1972555,Ritus1985,corrêa2020bosonicfermionicpropagatorsexternal} to calculate the inverse thermal propagator in momentum space. 
The Ritus method is a Fourier-like method that uses eigenfunction matrices $E_{p}^l(x)$ to diagonalize the inverse thermal propagator in momentum space. 
The eigenfunction matrices $E_p^l (x)$ are eigenvectors of the Dirac operator in a rigidly rotating reference frame. 
They are explicitly given as:
\begin{align}
    E_p^l (x) &= e^{i(p_zz+l\theta+\tau \omega_n)} \times \nonumber \\
    &\times\mathrm{diag}\left(J_{-}(\tilde{x}),J_{+}(\tilde{x}),J_{-}(\tilde{x}),J_{+}(\tilde{x})\right) \;,
\end{align}
where $J_{\pm}(\tilde{x}) \equiv J_{l\pm\frac{1}{2}}(p_\bot r)$ are the cylindrical Bessel functions of the first kind.
With this the inverse thermal propagator takes the form:
\begin{widetext}
    \begin{align}
    G_l^{-1}(p,p') &= \int d^4x d^4x' \, \bar{E_p}^l(x)
   (i\gamma_E^a D_{E,a} -\sigma -\gamma_E^3\gamma_E^5|\mathbf{s}|)
    \delta^{(4)} (x-x')  E_{p'}^{l'}(x') \nonumber \\
    & = (2\pi)^4 \delta^{(4)}(p-p') \delta_{l,l'} \tilde{G}_l^{-1} (\bar{p})\;,
\end{align}
\end{widetext}
where we have abbreviated the space-time integral 
\begin{equation}
\int d^4x\equiv \int d^3 x\int_0^\beta d\tau  \, 
\end{equation}
and where
\begin{align}\label{eq:inverse_propagator_momentum}
        \tilde{G}_l^{-1} (\bar{p}) = i\gamma_E^a \bar{p}_a -\sigma -\gamma_E^3\gamma_E^5|\mathbf{s}| \, 
\end{align}
is the momentum-space diagonalized inverse thermal propagator. 
To determine $\bar{p}_a$ we solve the equation:
\begin{align}\label{eq:eigenvalue_equation}
    \gamma_E^a D_{E,a} E_p^l (x) \overset{!}{=}  E_p^l (x) i \gamma_E^a \bar{p}_a \;, \end{align}
where the left-hand side is given as:
\begin{align}
    \gamma_E^a D_{E,a} E_p^l(x) = \begin{pmatrix}
    \Xi & \Phi \\
    -\Phi & -\Xi &
\end{pmatrix} \, .
\end{align}
 The matrix on the right-hand side is composed of $2 \times 2$-block matrices:
\begin{align}
    \Xi&= (-ip_0+l\Omega)\sigma_0 \mathcal{J}~,\\
    \Phi& = -\left(p_z\sigma_2 + p_\bot\sigma_3\right) \mathcal{J}~,\\
    \mathcal{J} &=e^{i(p_zz+l\theta+\tau \omega_n)}\left(J_{l-\frac{1}{2}}(p_\bot r),J_{l+\frac{1}{2}}(p_\bot r)\right)^T~,
\end{align}
where $\sigma_\mu$ are the Pauli matrices.
Solving \cref{eq:eigenvalue_equation} determines $\bar{p}_\mu$:
\begin{align}\label{eq:p_mu_bar}
    \bar{p}_{a} = (i(p_0 + il\Omega),0,ip_{\bot},ip_z)\; .
\end{align}
To obtain  \cref{eq:inverse_propagator_momentum} we used the orthogonality relation:
\begin{align}
    \int d^4x \bar{E_p}^l(x)\, E_{p'}^{l'}(x') = (2\pi)^4 \delta^{(4)}(p-p') \delta_{l,l'}\;.
\end{align}
Now we can write the effective potential \cref{eq:Veff_1} as:
\begin{align}
    V_{\text{eff}}\left(\sigma, |\mathbf{s}|\right) &= \frac{1}{2G} \left(\sigma - m \right) ^2 + \frac{1}{2 G_A} \mathbf{s}^2 \nonumber\\
    &+ \frac{i}{V} \Tr{\ln{(i\gamma_E^a\bar{p}_a-\sigma - \gamma_E^3\gamma_E^5|\mathbf{s}|)}}\ .
\end{align}
We perform the trace as far as possible and replace the integration in $\tau$ by a Matsubara sum:
\begin{align}\label{Eq:VeffwTrLog}
    V_{\text{eff}}\left(\sigma, |\mathbf{s}|\right) &= \frac{1}{2G} \left(\sigma - m \right) ^2 + \frac{1}{2 G_A} \mathbf{s}^2 \nonumber\\
    &+ \frac{iN_f N_c T}{(2\pi)^3V} \sum_{n,l=-\infty}^{\infty}\int p_\bot dp_\bot dp_z \nonumber \\
    &\times \Tr_D{\ln{(i\gamma_E^a\bar{p}_a-\sigma - \gamma_E^3\gamma_E^5|\mathbf{s}|)}} \, .
\end{align}
The above equation can be further simplified, leading to \cref{eq:Potential_final}, which will be used in the subsequent calculations throughout the paper.

\bibliography{magrot_bib}

\end{document}